\documentclass[10pt,aps,prl,twocolumn,letterpaper,nobibnotes,superscriptaddress,longbibliography,nofootinbib]{revtex4-2}
\usepackage{amsmath,amssymb,mathrsfs}
\usepackage{bm} 
\usepackage{natbib}
\usepackage{amsfonts}
\usepackage{rotating}
\usepackage{comment}
\usepackage{bbold} 
\usepackage{hhline}
\usepackage{color}
\usepackage{xcolor}
\usepackage{tabularray} 
\UseTblrLibrary{booktabs}
\usepackage[english]{babel}
\usepackage[utf8]{inputenc}
\DeclareUnicodeCharacter{0080}{}
\DeclareUnicodeCharacter{0093}{}
\usepackage[unicode=true,bookmarks=true,bookmarksnumbered=false,bookmarksopen=false,breaklinks=false,pdfborder={0 0 1},backref=false,colorlinks=true]{hyperref}

\hypersetup{linkcolor=magenta,urlcolor=blue,citecolor=blue,pdfstartview={FitH},hyperfootnotes=false,unicode=true}

\def\be{\begin{equation}}
	\def\ee{\end{equation}}
\def\bea{\begin{eqnarray}}
	\def\eea{\end{eqnarray}}
\def\nn{\nonumber}

\usepackage{algorithm}
\usepackage[noend]{algpseudocode} 

\begin{document}

\title{Hardware-Efficient Rydberg Atomic Quantum Solvers for NP Problems}

\author{Shuaifan Cao}
\affiliation{State Key Laboratory of Surface Physics, Institute of Nanoelectronics and Quantum Computing, and Department of Physics, Fudan University, Shanghai 200433, China}
\affiliation{Shanghai Qi Zhi Institute, AI Tower, Xuhui District, Shanghai 200232, China} 
\author{Xiaopeng Li} 
\email{xiaopeng\underline{ }li@fudan.edu.cn} 
\affiliation{State Key Laboratory of Surface Physics, Institute of Nanoelectronics and Quantum Computing, and Department of Physics, Fudan University, Shanghai 200433, China}
\affiliation{Shanghai Qi Zhi Institute, AI Tower, Xuhui District, Shanghai 200232, China} 
\affiliation{Shanghai Research Center for Quantum Sciences, Shanghai 201315, China} 
\affiliation{Hefei National Laboratory, Hefei 230088, China}
\date{\today}

\begin{abstract}
Developing hardware-efficient implementations of quantum algorithms is crucial in the NISQ era to achieve practical quantum advantage. 
In this work, we construct a quantum solver for NP problems based on Grover’s search algorithm, specifically tailored for Rydberg-atom quantum computing platforms.
We design the quantum oracles in the search algorithm using parallelizable quantum gates in Rydberg atom systems, 
yielding a unified framework for solving a broad class of NP problems with provable quadratic quantum speedup. 
We analyze the experimental resource requirements considering the dynamical qubit connectivity in the optical tweezer array. 
The required qubit number scales linearly with the problem size, representing a significant improvement over existing Rydberg-based  quantum-annealing approaches that incur quadratic overhead. 
These results provide a concrete roadmap for future experimental efforts towards algorithmic quantum speedup in NP problem solving using Rydberg atomic systems. 
Our construction indicates that atomic qubits offer favorable circuit depth scaling compared to quantum processors with fixed local connectivity.  

\end{abstract}

\maketitle



\paragraph*{Introduction.---}
Solving NP problems is of vital interest in computer science. It is not only of academic interest due to the central role of the NP-complete class~\cite{2002_Kitaev_Book}, but also of practical relevance as a wide range of real-world optimization tasks in logistics, scheduling, hardware verification, and artificial intelligence naturally map to NP problems. Canonical examples include the satisfiability (SAT) problem which directly encodes logical constraints~\cite{Biere_2009_SAT_intro, Abio_2014_SAT_intro}, the maximum independent set (MIS) problem which relates to complex social networks~\cite{Daliri_2025_MIS_intro},
and the exact cover problem (ECP) which arises in resource allocation~\cite{Knuth_2000_ECP_intro}. Constructing computationally efficient solvers for these problems is in high demand across industrial applications. 

With rapid developments of quantum computing in recent years, constructing quantum algorithms for NP problems has attracted significant attention~\cite{Shor, Grover, Farhi_2001_AQC_intro, Hen_2011_AQC_intro, Farhi_2014_QAOA, Wang_2018_QAOA_intro, Zhou_2020_QAOA_intro, Liu_2021_ML_intro, Yarkoni_2022_QA_intro, Nguyen_2023_QAOA_intro}. 
While the factorization problem can be solved efficiently by quantum circuits via Shor's algorithm~\cite{Shor}, the optimal quantum speedup for general NP-hard problems remains an open question. 
One potential route towards quantum advantage is through Grover's search algorithm~\cite{Grover, Boyer_1998_Grover, Zalka_1999_Grover, Long_2001_Grover, Brassard_2002_Grover, Toyama_2013_Grover, Yang_2022_SAT, Zhou_2024_Grover, Sun_2025_Grover}, which is expected to reduce
the computational complexity of finding a solution for NP problems involving 
$n$ binary variables from $2^n$ to $2^{n/2}$  in the worst case under the strong exponential-time hypothesis~\cite{2001_SETH,2001_kSAT}.   
This quadratic speedup holds considerable promise for industrial applications.
In the present NISQ era, it is essential to design hardware-efficient quantum solvers to achieve algorithmic quantum speedup for existing quantum devices with various limitations~\cite{2018_Preskill_NISQ}.

The past five years have witnessed remarkable progress in Rydberg atomic quantum computing. 
The number of individually controllable atomic qubits has increased rapidly—from tens~\cite{Bernien_2017_51qubit} to hundreds~\cite{Scholl_2021_Ryd, Ebadi_2021_Ryd, Bluvstein_2024_Ryd} and, more recently, to thousands~\cite{Manetsch_2024_array}—unlocking new opportunities for realizing algorithmic quantum speedup in solving NP problems. 
Remote entangling quantum gates have been realized with great parallelism~\cite{Evered_2023_gate} by dynamically moving atoms around through Acousto-Optical Deflectors (AOD), which defines the unique qubit connectivity of the system~\cite{Bao_2023_AOD, Adams_2024_AOD, Ferri_2022_AOD, Florshaim_2024_AOD}. 

On this platform, quantum annealing (QA) and quantum approximation optimization algorithms (QAOA)~\cite{2020_Lukin_PRX} have demonstrated intriguing empirical performance~\cite{Ebadi_2022_QA, Tibaldi_2025_QAOA,2022_Ahn_NatPhys, Graham_2022_circuit}. However, realizing quantum speedup with these approaches is challenging for two reasons: (1), their theoretical quantum speedup is not yet rigorously established; (2), the required optimal Hamiltonian encoding requires case-by-case design~\cite{Lucas_2014_AQC,2020_Li_PRXQ,2023_Lechner_PRL,2023_Pichler_PRXQ,Ye_2023_atom,2024_Schmelcher_PRR}. In contrast, Grover's search with its provable quadratic quantum speedup provides a promising alternative for building quantum solvers based on Rydberg atom arrays. 
Nevertheless, the construction of a gate-based Rydberg quantum solver for NP problems utilizing its unique non-local connectivity is lacking, and the minimal resource requirement for its implementation remains unknown.

\begin{table}[tp]
\centering
\begin{tblr}{
  colspec = { 
    Q[c, m, wd=2cm]   
    Q[c, m, wd=0.17cm] 
    Q[c, m, wd=0.17cm] 
    Q[c, m, wd=0.17cm] 
    Q[c, m, wd=1cm] 
    Q[c, m, wd=2.6cm]   
  },
  row{1} = {rowsep=1pt, abovesep = 3pt, belowsep = 2pt}, 
  row{2} = {rowsep=0pt, abovesep = 2.5pt, belowsep = 0pt},
  row{3-5} = {rowsep=0pt, abovesep = 0pt, belowsep = 0pt}, 
  row{6} = {rowsep=0pt, abovesep = 0pt, belowsep = 1pt}
}
\toprule
 Oracle & $\tilde{n}$ & $N$ & $t$ & Qubits & Circuit depth \\ 
\midrule
 $(n, m, k)$ SAT & $n$ & $m$ & $k$ & \SetCell[r=5]{m} $n+2N$ & \SetCell[r=5]{m} $O(\mathrm{polylog}(N))$ \\ 
 $(n, v, d)$ SCP & $n$ & $v$ & $d$ & & \\ 
 $(n, v, d)$ ECP & $n$ & $v$ & $d$ & & \\ 
 $(n, e)$ MIS   & $n$ & $e$ & $2$ & & \\ 
 $(n, e)$ MCP   & $n$ & $e$ & $2$ & & \\ 
\midrule
 \SetCell[c=4]{c} {Grover at $N=O(n)$:} & & & & $O(n)$ & $O(\mathrm{polylog}(n)2^{n/2})$ \\
\bottomrule
\end{tblr}
\caption{\label{tab:Results} Cost of five representative NP problems using the Rydberg atomic quantum solver. 
The parameter terminology follows the standard as in Refs.~\onlinecite{Karp:NPC, Garey_1990_NPC}. The number of binary variables is $n$.  
For $k$-SAT, $m$ ($k$) is the number of clauses (variables in each clause); for MIS/MCP, $e$ is the number of edges in the graph; for $d$-regular SCP/ECP~\cite{Moore_2015_d_regular}, $v$ ($d$) is the number of elements (subsets that contain one chosen element). 
The parameters of the solver, 
$(\tilde{n},N,t)$ are the number of (data qubits, checking units (Fig.~\ref{fig:Remapping}), data qubits needed in each checking unit). The hardness thresholds of these problems appear at $N=O(n)$~\cite{MCP_threshold,SAT_threshold, Moore_2015_d_regular, 3ECP_threshold, Weight_2001_VCP_threshold, Coja_2015_MIS_threshold}. 
For $N=O(n^2)$, the qubit number cost remains linear using a variant encoding scheme (Supplementary Materials). 
}
\end{table}

In this Letter, we develop a generic Rydberg atomic quantum solver based on Grover's search algorithm, applicable to a broad class of NP problems, including all Karp's 21 NP-complete problems~\cite{Karp:NPC}. 
This is achieved by constructing Grover search oracles for different NP problems through parallelizable single-qubit and Rydberg CZ/CCZ gates, incorporating the unique qubit connectivity enabled by AOD in atom tweezer arrays~\cite{Polimeno_2018_light_tweezer, Kaufman_2021_light_tweezer, Volpe_2023_light_tweezer}. 
As illustrative examples, we analyze five standard NPC problems, including SAT, MIS, ECP, the max-cut problem (MCP), and the set cover problem (SCP). As summarized in Tab.~\ref{tab:Results}, the qubit number cost follows a linear scaling with the problem size, in contrast to the quadratic scaling inherent in previous Rydberg-based encodings for QA and QAOA solvers~\cite{2020_Li_PRXQ,2023_Lechner_PRL,2023_Pichler_PRXQ,Ye_2023_atom}. 
The circuit depth scales as $\operatorname{polylog} (n)\, 2^{n/2}$, and an additional  $O(\sqrt{n} \log n \, 2^{n/2})$ classical atomic transports are needed (almost perfect fidelity), yielding a significant speedup over the classical computation complexity of $2^n$ in the worst case. 
These results provide concrete experimental protocols and resource requirements that advance Rydberg-atom quantum computing toward quantum advantage in solving NP problems. 

\paragraph*{A unified framework for Grover's search-based NP quantum solver.---}
To apply Grover's search to NP problems, we need to construct an NP-oracle that produces a $\pi$-phase shift for the legal (solution) states with respect to the illegal (non-solution) states~\cite{Grover}. 
The decision version of an NP problem with $n$ binary variables $z_i\in \{0, 1\},i=1,\ldots, n$ is characterized by a binary function $f(z)$, where $z=(z_1, \ldots, z_n)$ and $f(z^*)=1$ indicates a solution $z^*$. 
We find that for a broad class of NP problems, $f(z)$ has a more refined structure, yielding a unified decomposition: 
\begin{equation}
    f(z)=\bigwedge_{\mu=1}^{p} g_\mu({\cal A}_\mu) \wedge H\left(\sum_{\nu=1}^{q} h_\nu({\cal B}_\nu)-k_1\right),  \label{eq:f(x)}
\end{equation}
where notations are explained as follows. The functions $g_\mu$ represent $p$ logical constraints, each involving a subset of variables ${\cal A}_\mu =\{z_i| i \in {\cal I}^A_\mu\}$ where ${\cal I}^A_\mu=\{i_{\mu 1}, i_{\mu 2}, \ldots \}$ denotes the index set of variables involved in the $\mu$-th constraint. 
We also incorporate an inequality constraint by $\textstyle H\left(\sum_{\nu} h_\nu({\cal B}_\nu)-k_1\right)$,  
where $H(x)$ is the Heaviside step function, $k_1$ is a given threshold, $h_\nu$ are integer-valued functions, and ${\cal B}_{\nu=1,\ldots ,q} = \{z_i| i\in{\cal I}_\nu^B\}  $ are subsets of variables with index sets ${\cal I}^B_\nu$. 
Such constraint naturally arises in problems such as Knapsack, MCP and MIS~\cite{Karp:NPC}. 
Here $k_1$, $g_\mu$, and $h_\nu$ are determined by the specific NP problem instance. 



\begin{figure}[tp]
    \begin{center}
  \includegraphics[width=1\linewidth]{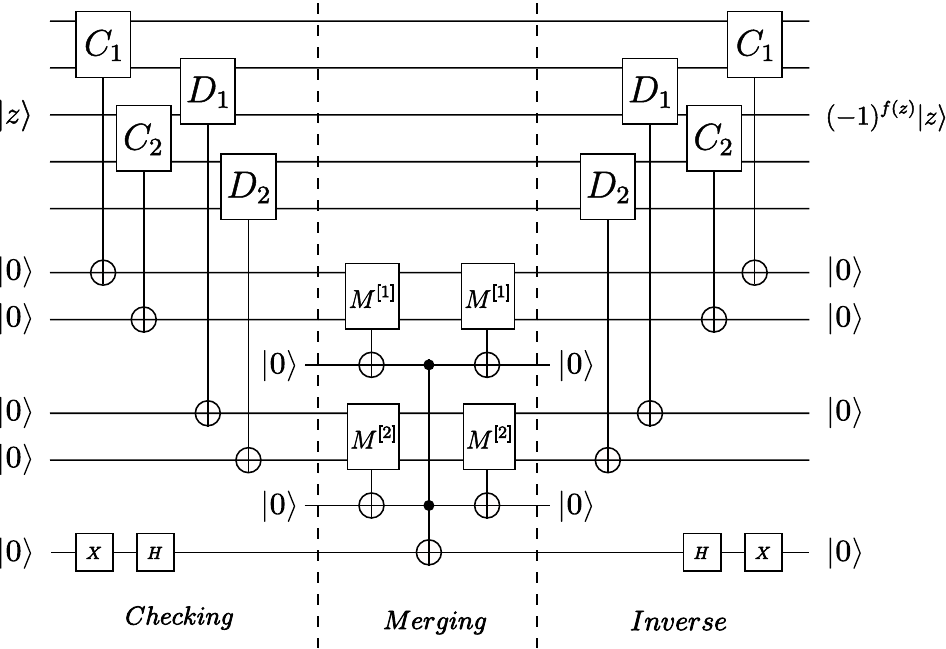}
    \end{center}
  \caption{\label{fig:Unified} 
  Schematic of the NP-oracle circuit (Eq.~\ref{eq:oracle}). Data qubits $|z\rangle$ encode the problem variables and the rest serve as ancillae. The circuit comprises checking units ($C_\mu, D_\nu$) and merging blocks ($M^{[1]}, M^{[2]}$), as detailed in the {\it main text}. For simplicity, the parameter $b$ ({\it main text}) is set to 1. 
  }
\end{figure}

We introduce two types of checking units, $C_\mu$, and $D_\nu$, which are unitary operations defined by 
\be 
    \begin{cases}
        C_\mu|{\cal A}_\mu\rangle|0\rangle \mapsto |{\cal A}_\mu\rangle|g_\mu({\cal A}_\mu)\rangle, \\
        D_{\nu}|{\cal B}_\nu \rangle|0\rangle^{\otimes b} \mapsto |{\cal B} _\nu\rangle|h_\nu ({\cal B}_\nu )\rangle. 
    \end{cases}
\ee 
A certain number ($b$) of ancillae is needed to represent the integer-valued function $h_\nu$. 
Generating the $\pi$-phase shift in Grover's oracle requires merging the information stored in the ancillae. 
To implement that, we define two unitary merging blocks $M^{[1]}, M^{[2]}$ as: 
\bea 
  &&  M ^{[1]} \left( \otimes_\mu |g_\mu({\cal A}_\mu)\rangle \right) |0\rangle = \left( \otimes_\mu|g_\mu({\cal A}_\mu )\rangle \right) |\bigwedge_\mu  g_\mu({\cal A} _\mu)\rangle, \\
  &&  M^{[2]} \left( \otimes_\nu|h_\nu({\cal B}_\nu)\rangle \right) |0\rangle = \left(\otimes_\nu|h_\nu({\cal B}_\nu)\rangle \right) |H (\sum_\nu h_\nu({\cal B}_\nu)-k_1 )\rangle.  \nn \label{eq:M1M2}
\eea 

With the building blocks introduced above, we construct a Grover search oracle for NP problems, consisting of three steps (Fig.~\ref{fig:Unified}), 
\begin{align}
    &|z\rangle |0\rangle^{\otimes \left(p+bq\right)}|-\rangle \notag \\
    &\xrightarrow[\mathrm{Check}]{C_\mu, D_\nu} |z\rangle 
    \left( \otimes_\mu |g_\mu({\cal A}_\mu)\rangle \otimes_\nu|h_\nu({\cal B} _\nu)\rangle \right) |-\rangle 
    \notag \\
    &\xrightarrow[\mathrm{Merge}]{M^{[1]},M^{[2]}} |z\rangle \left (\otimes_\mu|g_\mu({\cal A}_\mu)\rangle\otimes_\nu|h_\nu({\cal B}_\nu)\rangle \right) \left( (-1)^{f(z)}|-\rangle \right) \notag \\ &\xrightarrow[\mathrm{Inverse}]{} 
    \left( (-1)^{f(z)}|z\rangle \right) |0\rangle^{\otimes \left(p+bq\right)}|-\rangle, \label{eq:oracle}
\end{align}
where $|z\rangle$ are data qubits that encode the variables and the rest are ancillae. 
For standard NP problems, it is typical that $g_\mu$ ($h_\nu$) only contains a constant number of variables. In such cases, we find that the checking and merging circuits have efficient implementation using Rydberg tweezer arrays with their dynamically configurable qubit connectivity, as we describe in detail below.

\paragraph*{Rydberg quantum circuit for checking.---}
The construction of the checking units $C_\mu$ and $D_\nu$ for different NP problems is straightforward, as they consist of basic operations (Supplementary Materials). 
What it requires to develop a hardware-efficient solver is to utilize the unique long-range qubit connectivity of the Rydberg atom quantum computing system~\cite{Henriet_2020_Ryd_review, Bluvstein_2022_Ryd}, and implement the units with maximal parallelism. 

In the Rydberg system, the qubit connectivity and gate-level parallelism are determined by crossed Acousto-Optical Deflectors ($\times$AOD)~\cite{Trypogeorgos_2013_AOD, Chisholm_2018_AOD, Bluvstein_2022_Ryd, Florshaim_2024_AOD}, which produce dynamical tweezer beams carrying atoms around individually. 
Due to the one-dimensional nature of the acousto-optical deflection, the atom array transported by $\times$AOD forms a tensor grid, denoted as $\mathcal{G}= R \times C \equiv \{(x,y)|x\in R, y\in C\}$, where $R$ and $C$ are rows and columns of the tensor grid. 
The two-qubit gates between atoms from two tensor grids ${\cal G}$ and ${\cal G}'$ (=$R'\times C'$) correspond to a map $F: \mathcal{G} \mapsto \mathcal{G}'$. 
By moving atoms around with $\times$AOD, these two-qubit gates can be performed in parallel if $F$ 
has a product-form, 
\be
 F = f_r \times f_c, 
 \label{eq:CZ}
\ee
where $f_r: R \mapsto R'$ and $f_c: C \mapsto C'$ are monotonically increasing bijections.  
This defines a tensor-grid qubit connectivity, which is unique to the present Rydberg system. 
Such parallel two-qubit gates have been realized in the Rydberg system with high fidelity~\cite{Levine_2019_gate, Evered_2023_gate}.  

A crucial mathematical structure in this framework for NP problems (Eq.~\eqref{eq:f(x)}) is that the circuit realizations of $C_\mu$s (and likewise for $D_\nu$s) become identical up to some instance-specific single-qubit gates, once the NP problem is specified---such as $k$-SAT, MIS, MCP, $d$-SCP, or $d$-ECP---enabling highly parallel implementation with Rydberg atoms. 
The resource requirement of the Rydberg quantum solver is characterized by three key parameters: the number of data qubits $\tilde{n}$, the number of checking units $N$, and the number of data qubits involved in one checking unit $t=|\mathcal{A}_{\mu}|,\forall \mu$.
Their dependence on the problem size is provided in Tab.~\ref{tab:Results}. 

\begin{figure}[btp]
  \includegraphics[width=\linewidth]{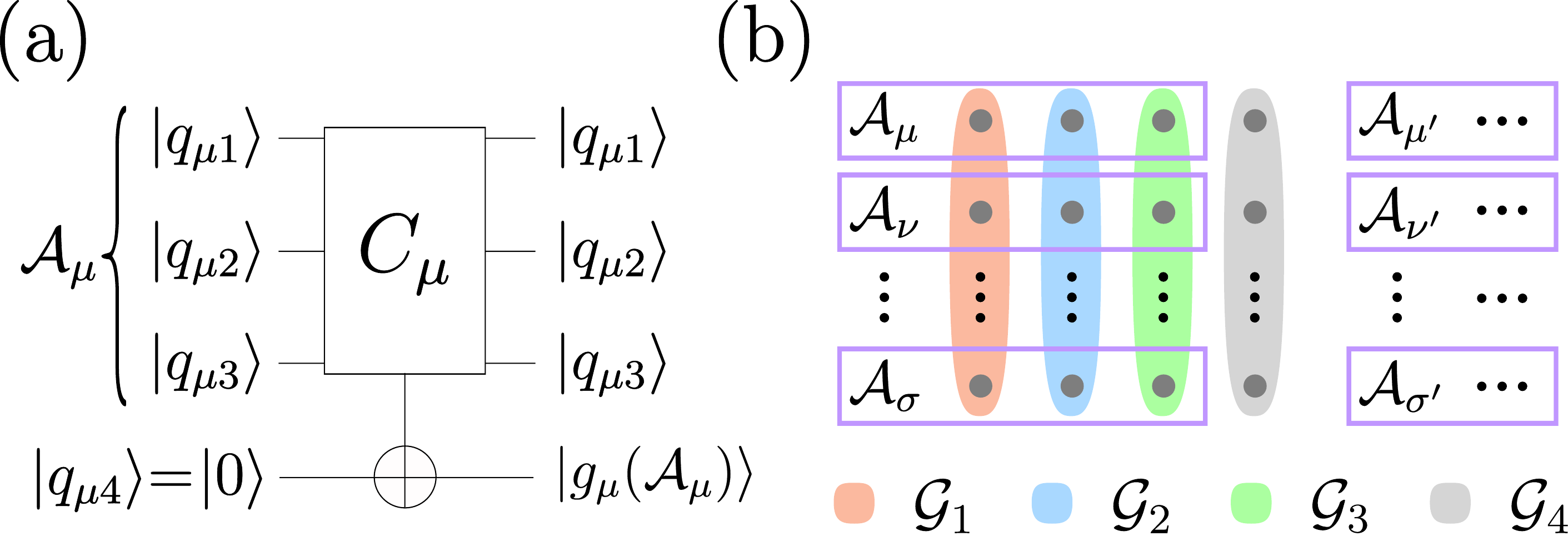}
  \caption{\label{fig:Remapping} 
Parallelization of checking units ($t=3$) by properly arranging atoms to satisfy the qubit mapping (Eq.~\ref{eq:Mapping}). 
(a), the checking unit $C_\mu$. (b), the qubit mapping. 
Collect the qubit at a certain position in each checking unit together and they form a tensor grid. 
Two-qubit gates between tensor grids $\mathcal{G}_\tau =\bigcup_{\mu} P(q_{\mu \tau})$ are parallelizable, indicating the parallelism of the checking units. }
\end{figure}

Now, we provide a protocol to implement multiple checking units in parallel. 
We take the parallel realization of $C_\mu$ as an illustrative example. The procedure remains the same for $D_\nu$. Taking $O$ checking units, ($C_{\mu_1}, C_{\mu_2}, \ldots, C_{\mu_O}$), with their indices forming a set ${\cal O}$, we assume a minimal requirement: these different checking units do not share overlapping qubits, i.e.,  
\be 
{\cal A}_{\mu} \cap {\cal A}_{\mu'} = \emptyset\,\, {\rm for} \,\, \mu,\mu' \in {\cal O}.  
\label{eq:nocollision}
\ee 
This assumption is minimal because parallel implementation of multiple two-qubit gates sharing overlapping qubits is fundamentally impossible. 
The qubits involved in $C_\mu$ are $q_{\mu\tau}$, with $\tau = 1, \ldots, t+1$.  
For parallelization, we rearrange the qubits to the physical atomic positions, $[x,y]$, according to 
\begin{equation}
q_{\mu \tau} \to [x(q_{\mu \tau} ), y(q_{\mu \tau} ) ] \equiv P(q_{\mu \tau}), \notag
\end{equation}
with 
\begin{equation}
\label{eq:Mapping}
P(q_{\mu \tau}) = \left[ (t+1)\left \lfloor (\mu-1)/ \lceil \sqrt{n} \rceil \right \rfloor+\tau-1, (\mu-1)\ \mathrm{mod}\ \lceil \sqrt{n}\rceil\right] . 
\end{equation} 
We emphasize three key properties of this map (Fig.~\ref{fig:Remapping})---(i): It maps different qubits to different atom positions; 
(ii):  Collecting all $\mu\in {\cal O}$ with a fixed $\tau$, the mapped atom positions, $P(q_{\mu \tau})$, form a tensor grid ${\cal G}_\tau$ (after full-filling the last column with some completing atoms); 
(iii): Taking two tensor grids with  $\tau \neq \tau'$, ${\cal G}_\tau$ and ${\cal G}_{\tau'} $ are related to each other by a simple translation, for  
$P(q_{\mu \tau})-P(q_{\mu \tau'}) =(r_{\tau \tau'}, c_{\tau \tau'})$ is independent of $\mu$.
The atom rearrangement of the qubits involved in the $O$ checking units 
corresponding to $P$ requires at most $tO$ atomic transports with $\times$AOD. 

In this map, the two-qubit gates between $q_{\mu \tau_1}$ and $q_{\mu \tau_2}$ for all $\mu \in {\cal O}$ can be performed in parallel, since they correspond to a map 
$F_{\tau \tau'} : {\cal G}_{\tau_1} \mapsto {\cal G}_{\tau_2}$, which is given by 
$F_{\tau \tau'}=f_1 \times f_2$,  
where $f_1(x)=x+r_{\tau \tau'}, f_{2}(y)=y+c_{\tau \tau'}$. 
The single-qubit gates on a tensor grid are also parallelizable (Supplementary Materials).

With the above protocol, the circuit realization of the checking units boils down to separating all ${\cal A}_\mu$s into a series of subgroups, (${\cal O}_1, {\cal O}_2, \ldots, {\cal O}_L$). Each of the $L$ subgroups satisfies the condition in Eq.~\eqref{eq:nocollision}. Finding these subgroups can be solved by mapping the problem to maximal matching on hypergraphs. We consider a hypergraph $G(V,E)$, with $n$ binary variables as its vertices, and $\mathcal{A}_\mu$ as its edges, i.e., $E=\{\mathcal{A}_\mu |\mu = 1,\ldots,p \}$. The corresponding algorithm is described as follows: 
\newpage
\begin{center}
\textbf{Algorithm 1.} 
\end{center}
\begin{center}
\begin{minipage}{0.48\textwidth}
\begin{algorithmic}[1]
\State $G_1(V_1, E_1) \gets G(V, E)$
\State \texttt{Mlist} $\gets$ [ ]
\While{$E_1 \neq \emptyset$}
    \State Find a maximal matching $\texttt{M}$ in $G_1$
    \State $G_1 \gets G_1\setminus \texttt{M}$; append \texttt{M} to \texttt{Mlist}
\EndWhile
\State return \texttt{Mlist}
\end{algorithmic}
\end{minipage}
\end{center}
The subgroups of ${\cal O}$s are obtained sequentially following this algorithm. 
The step of finding a maximal matching can be performed classically and efficiently using Edmonds' Blossom Algorithm~\cite{EBA} 
(for $t=2$ where $G$ reduces to a graph) 
and greedy algorithms~\cite{Besser_2015_greedy} (for $t>2$). 
Algorithm 1 is executed only once ahead of the Grover iterations and its complexity is negligible.

{ 
The complete checking process consists of $L$ layers of parallel Rydberg quantum gates, 
and at most $O(L\sqrt{n} \log n)$ classical atomic transports implemented via AOD-based rearrangement. 
Numerical results indicate that $L$ is of the order, $O(N/n)$ for a random problem instance (Supplementary Materials).
Since one checking unit typically has constant depth, 
the total depth of the checking circuit scales as $O(N/n)$, which reduces to $O(1)$ for $N=O(n)$. 
Compared to quantum swap gates, atomic transport is advantageous due to its significantly higher fidelity~\cite{Bluvstein_2024_Ryd}. 
The cost of such classical operations is not included in the circuit depth reported in Tab.~\ref{tab:Results}. 
} 


\begin{figure}[tp]
 \includegraphics[width=1\linewidth]{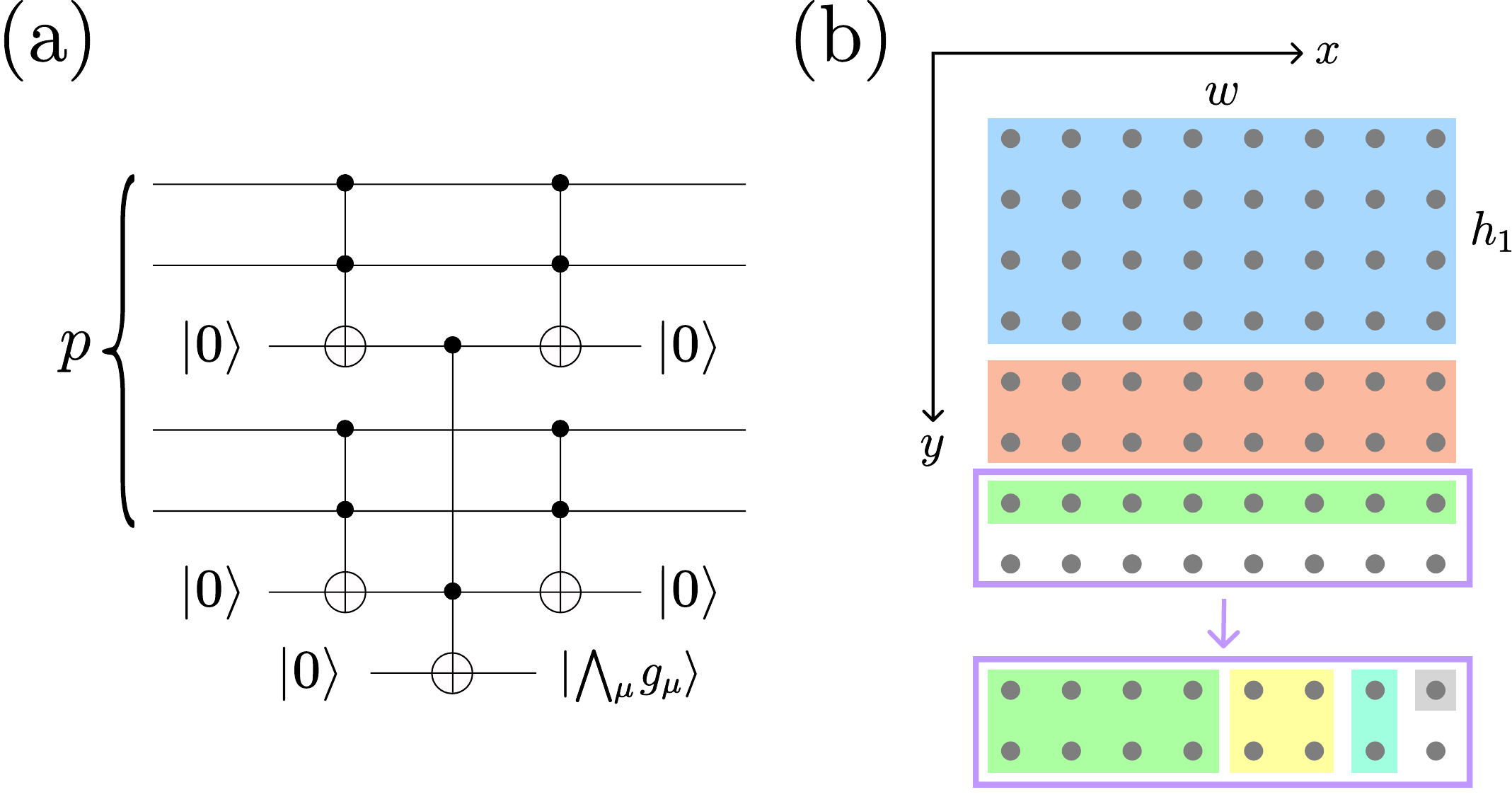}
  \caption{\label{fig:Merging} Implementation of the merging block $M^{[1]}$. 
  { 
  (a), the circuit for $M^{[1]}$ (QBT), which supports recursive extension. (b), transpilation of the circuit onto the Rydberg-atom array.} 
  The blue region contains the output of the checking circuit to be merged. 
  The information is merged from up to down (see {\it main text}). 
  Atoms in different regions represent qubits at different levels of QBT. 
  One rearrangement is used in the purple frame to make the total region more compact. }
\end{figure}

\paragraph*{Rydberg quantum circuit for merging.---}
To implement the merging operations $M^{[1]}$ and $M^{[2]}$, we present a highly parallelizable circuit construction tailored to the tensor-grid qubit connectivity of the Rydberg-atom quantum computing systems. 
A direct realization of $\bigwedge_\mu  g_\mu$, and $H (\sum_\nu h_\nu({\cal B}_\nu)) $, as required by $M^{[1]}$ and $M^{[2]}$ generally involves $O(N)$ gates. 
These gates are typically non-local, and on quantum processors with local connectivity---such as standard superconducting qubit~\cite{Linke_2017_sc_local, Kjaergaard_2020_sc_local} or quantum dot architectures~\cite{Qiao_2020_dot_local, Burkard_2023_dot_local}---the corresponding circuit depth necessarily scales as $O(\mathrm{poly}(N))$~\cite{Chu_2023_sc_gate}. 
In contrast, we show below that the same merging operations can be implemented with a circuit depth scaling as $O({\rm polylog}(N))$ on Rydberg-atom arrays, exploiting their dynamically reconfigurable qubit connectivity.  This leads to an exponential advantage in depth over conventional quantum computing architectures with fixed local connectivity. 

For simplicity, we focus on the parallel implementation of the merging operation of $M^{[1]}$. 
Following the checking circuit, 
{ the results of $g_\mu$ are assumed to be stored in ancilla qubits arranged in a rectangular grid with height $h_1$, width $w$, and $h_1 w =p$,}   
labeled by their positions as $|g_{xy}\rangle$, where $x\in[0,w-1]$, and $y\in [0, h_1-1]$. To implement merging, we introduce an additional layer of ancillae located at positions $x\in [0,w-1]$, $y\in [h_1,3h_1/2-1]$ (Fig.~\ref{fig:Merging}). We apply Toffoli (CCX) gates, 
\be 
{\rm CCX} |g \rangle |g'\rangle |0 \rangle 
= |g \rangle |g' \rangle  |g  \wedge g' \rangle,
\ee 
for qubit triplets located at 
$(x,y)$, $(x, y+1)$, $(x, h_1+y/2)$, for all even $y$. 
These gates can be constructed with the Rydberg blockade mechanism, which can be performed simultaneously by shining global Rydberg lasers on the atoms~\cite{Levine_2019_gate, Evered_2023_gate}. 
The introduced layer of ancillae forms a new rectangular grid with a reduced height $h_2=h_1/2$, and the width remaining the same. 
The above procedure is then iterated until the height is reduced to $1$. 
To finalize the merging, the leftover single-line of ancillae is split into two halves, and rearranged into  a tensor grid with a height, $2$ (Fig.~\ref{fig:Merging}). The iteration procedure then continues  until $\bigwedge_\mu  g_\mu$ is obtained. 
Finally, the ancilla qubits introduced during the process need to be restored. 
All these operations are compatible with the $\times$AOD techniques~\cite{Trypogeorgos_2013_AOD, Chisholm_2018_AOD, Bluvstein_2022_Ryd, Florshaim_2024_AOD}. 

\begin{figure}[tp]
    \includegraphics[width=0.75\linewidth]{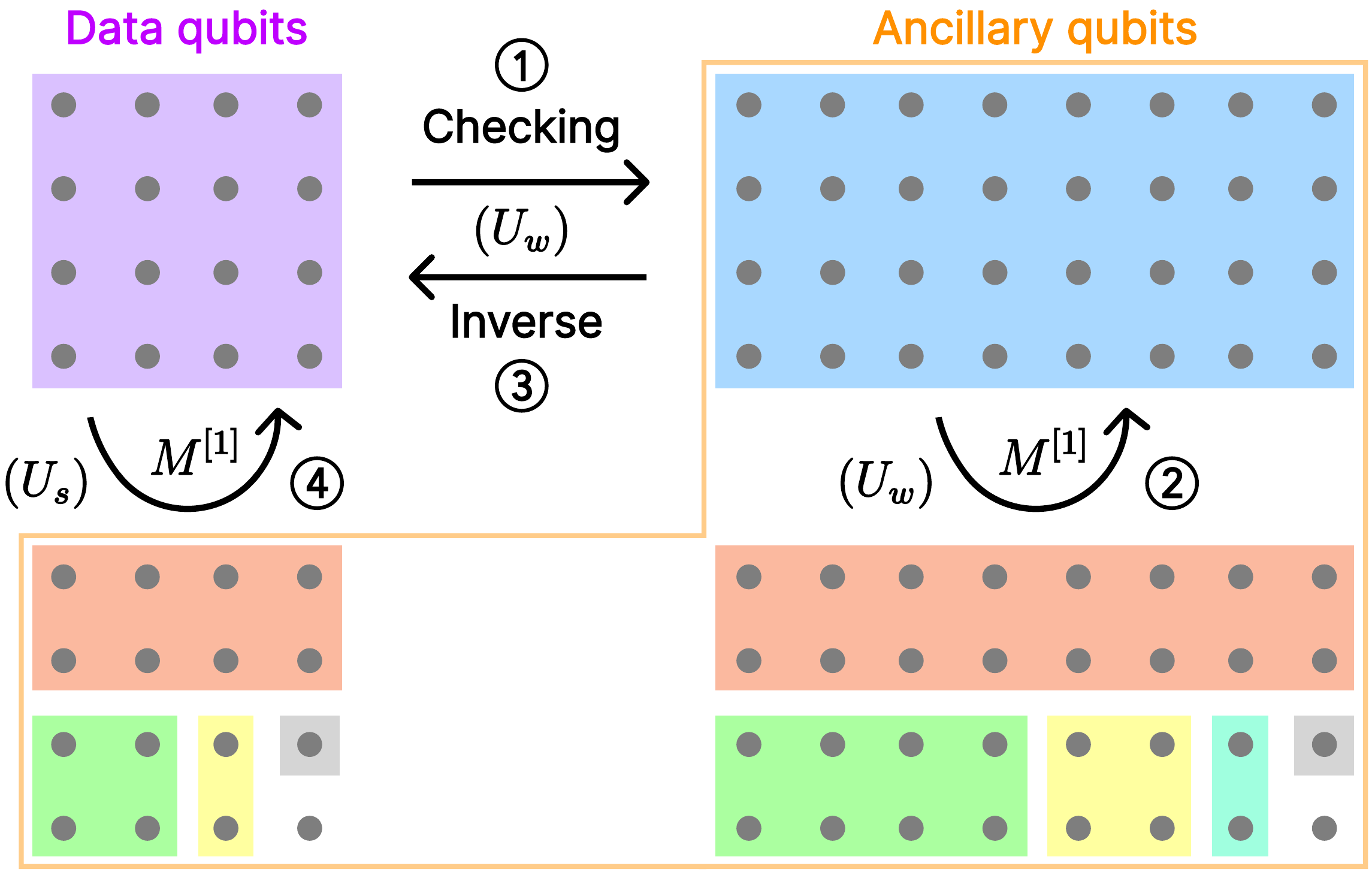}
  \caption{
  { Schematic of one Grover iteration in the Rydberg quantum solver. The Grover oracle ($U_\omega$) is constructed using checking and merging circuits ($M^{[1]}$ as an example) in step 1--3.
  The Grover diffusion operator ($U_s$) has a straightforward implementation using $M^{[1]}$ (Supplementary Materials).
  The curved arrow associated with $M^{[1]}$ indicates the restoration of ancillae, which is necessary to induce phase kickback.
  } 
}
  \label{fig:mainSolver}
\end{figure} 

In the entire merging process, the binaries $g_\mu$ are merged in a pairwise manner, forming a binary tree structure, dubbed quantum binary tree (QBT). 
The operations in $M^{[2]}$ can also be implemented following a similar iteration approach. We also construct a quantum recursive adder (QRA) to realize $M^{[2]}$ 
{ (Supplementary Materials).}  
{ 
Assuming that the checking circuit yields $N$ outputs, the total number of ancilla qubits required for QBT and QRA is $2N$ and $(b+1)N$, respectively. 
The corresponding circuit depths scale as $O(\mathrm{log}_2N)$ for QBT, and $O((\mathrm{log}_2N)^2)$ for QRA, leading to the resource requirements summarized in Table~\ref{tab:Results}.
The Rydberg circuit depths for QBT and QRA exhibit favorable scaling compared to implementations on quantum processors with fixed local connectivity, where a circuit depth exceeding $O(\sqrt{N})$
 is required (Supplementary Materials)~\cite{Chu_2023_sc_gate}. 
}


With the construction of checking and merging circuits, the Grover oracle for solving NP problems (Fig.~\ref{fig:Unified}) is completed. 
The corresponding Rydberg atomic solver for NP problems is illustrated in Fig.~\ref{fig:mainSolver} and the cost in Tab.~\ref{tab:Results}. The qubit number cost is linear, which has a scaling advantage over the quadratic cost in previous quantum annealing algorithms~\cite{2020_Li_PRXQ,2023_Lechner_PRL,2023_Pichler_PRXQ,Ye_2023_atom}. 
The other problem with quantum annealing is that the time cost is difficult to bound for NP problems. 
In contrast, the circuit depth of our gate-based Rydberg solver has a definite scaling $O(\operatorname{polylog}(n)\,2^{n/2})$, providing a rigorous quadratic speedup over classical computing, assuming the strong exponential time hypothesis~\cite{2001_SETH,2001_kSAT}. 

\paragraph*{Discussion.---}
To conclude, we have developed a Rydberg atomic quantum solver for a broad class of NP problems, based on Grover search algorithm. 
Our protocol exploits the tensor-grid qubit connectivity inherent to Rydberg systems to realize a hardware-efficient implementation of the Grover oracle.
As concrete examples, we analyze the resource requirements for five representative NP problems as shown in Tab.~\ref{tab:Results}. 
These results provide a concrete experimental blueprint that advances Rydberg-atom quantum computing toward demonstrations of quantum advantage in solving NP problems. 
We remark that the scaling of the circuit depth of the quantum solver largely relies on the unique non-local connectivity of Rydberg atom systems. The implementation on other quantum processors with fixed local connectivity such as superconducting qubits is expected to have an additional polynomial overhead. 
Moreover, since this work provides a hardware-efficient realization of Grover oracles, the construction is also useful in other oracle-based algorithms such as quantum random walks~\cite{Montanaro_2016_random_walks, Zhou_2025_random_walks}. 

\paragraph*{Acknowledgements.---}
We acknowledge helpful discussion with Yueyang Min, Yingzhou Li and Shaoming Fei. 
This work is supported by  the Innovation Program for Quantum Science and Technology of China (Grant No. 2024ZD0300100),
the National Basic Research Program of China (Grants No. 2021YFA1400900), 
Shanghai Municipal Science and Technology (Grant No. 25TQ003, 2019SHZDZX01, 24DP2600100). 

\bibliographystyle{apsrev4-1}
%

\newpage
\renewcommand{\theequation}{S\arabic{equation}}
\renewcommand{\thesection}{S-\arabic{section}}
\renewcommand{\thefigure}{S\arabic{figure}}
\renewcommand{\thetable}{S\arabic{table}}
\setcounter{equation}{0}
\setcounter{figure}{0}
\setcounter{table}{0}

\begin{widetext} 

\newpage 

\begin{center}
    \Huge Supplementary Materials
\end{center}

In this Supplementary Material, we provide additional details complementing the main text of \textbf{Hardware-Efficient Rydberg Atomic Quantum Solvers for NP Problems}. The content is organized as follows. 
Section~.I reviews the qubit connectivity and the gate-level parallelism of the Rydberg atom system. 
Section~.II reviews the Grover’s search algorithm that the solver is based on. 
Section~.III provides detailed information on the checking circuit in the unified framework and its transpilation. 
Section~.IV provides detailed information on the merging circuit in the unified framework and its transpilation. 
Section~.V discusses the compatibility of our algorithm with some quantum error correction codes. 
Section~.VI estimates the precise cost of one Grover oracle for small problem sizes. 
Section~.VII compares the cost of realizing this framework in a Rydberg atom system and on a standard superconducting chip. 

\section{\label{sec:Ryd}I.\quad Rydberg atom system}
Rydberg atom systems~\cite{Urban_2009_Ryd, Ebadi_2021_Ryd, Scholl_2021_Ryd, Kaufman_2021_Ryd, Saffman_2024_Ryd, Bluvstein_2022_Ryd, Bluvstein_2024_Ryd, Wintersperger_2023_Ryd, Ebadi_2024_Ryd} have emerged in recent years as one of the most promising platforms for quantum computing. Utilizing the Rydberg blockade mechanism~\cite{Urban_2009_Ryd, Isenhower_2011_gate}, atoms exhibit strong and controllable interactions when brought close together and excited to high-lying Rydberg states. This enables entanglement between individually trapped neutral atoms and facilitates the implementation of high-fidelity two-qubit gates, which are essential for universal quantum computation. 
The system has significant advantages~\cite{Wu_2021_Rydreview, Shi_2022_Rydreview} compared to other platforms for quantum computing, including high-fidelity quantum gates~\cite{Isenhower_2010_gate, Evered_2023_gate, Jandura_2022_gate, Ma_2022_gate}, scalable and controllable large-scale atom array~\cite{Manetsch_2024_array, Anand_2024_array, Singh_2022_array} and long coherence time~\cite{Picken_2019_life, Ma_2022_gate, Chang_2025_life}. 
In particular, the advancement of optical tweezers~\cite{Ashkin_1999_light_tweezer, Kaufman_2021_light_tweezer, Polimeno_2018_light_tweezer, Volpe_2023_light_tweezer} allows for precise manipulation and transport of atoms and a high degree of parallelism~\cite{Evered_2023_gate, Levine_2019_gate}. 

In Rydberg atom systems, the precise manipulation of atoms is realized by the Acousto-Optical Deflector (AOD)~\cite{Trypogeorgos_2013_AOD, Chisholm_2018_AOD, Bluvstein_2022_Ryd, Ferri_2022_AOD, Adams_2024_AOD, Florshaim_2024_AOD} based optical tweezers. 
In experiments, using the crossed AOD ($\times$AOD), a beam can be deflected to one spot and then stepwise scans a 2D tensor grid pattern, forming a group of optical tweezers to fetch atoms and move them. 
Therefore, a group of atoms that form a tensor grid can be moved or illuminated by a laser simultaneously. If a group of atoms can be moved towards another group in parallel, identical one-to-one entangling gates between them can be performed in parallel (Fig.~\ref{fig:Tensor}). A rigorous description can be seen in the main text. 
This indicates the tensor-grid qubit connectivity of this system and the feasibility of implementing entangling gates between distant atoms. 
For single-qubit gates, their implementation also requires AOD to do atomic addressing and then shine lasers. Therefore, identical single-qubit gates targeted at atoms on a tensor grid can be performed in parallel. 
High-fidelity single-qubit gates and parallel entangling gates have been realized in the Rydberg atom system~\cite{Jandura_2022_gate, Levine_2019_gate, Evered_2023_gate}. 
\begin{figure}[htb]
  \includegraphics[width=0.4\textwidth]{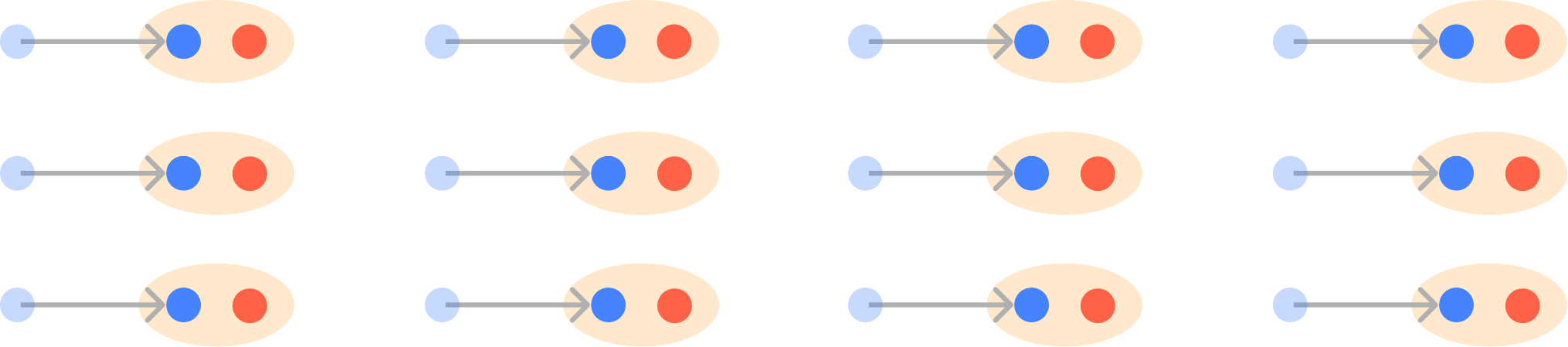}
  \caption{\label{fig:Tensor} Using $\times$AOD to move atoms and perform parallel entangling gates. The blue group of atoms can be moved towards the red group in parallel. The Rydberg lasers (orange) can also be shined in parallel. For CCZs, one more tensor grid of atoms needs to be moved into proximity.  }
\end{figure}

In this system, atoms can be classically transported by $\times$AOD, instead of using swap gates as that in superconducting with fixed local connectivity. 
The time to move an atom across $n$ sites is $O(n^{1/2})$~\cite{Zhang_2015_movetime, Bluvstein_2022_Ryd, Endres_2016_rearrange}, as $\times$AOD has an upper bound in acceleration. 
Assuming that the number of variables is $n$, their corresponding qubits are mapped to a 2D atom array with side lengths of order $\sqrt{n}$. 
Therefore, the average time to relocate one atom in this array is $O(n^{1/4})$. 
{
Importantly, the fidelity of atomic transport with $\times$AOD is much higher than that of gate operations (almost perfect)~\cite{Evered_2023_gate, Bluvstein_2024_Ryd, Chiu_2025_move}, even for long distance. Therefore, we do not include the atomic transports in the circuit depth but leave it to the classical depth for their robustness. }

The state-of-the-art fidelity of quantum gates in Rydberg atom systems is shown in the following. 
Basic Rydberg gates that can be performed directly in this system include arbitrary single-qubit gates, CZ and CCZ, with respective fidelity of $99.9\%, 99.7\%,97.9\%$ on the state-of-the-art platform~\cite{Evered_2023_gate, Bluvstein_2024_Ryd}. The entangling gates can be performed between any atom using the AOD technique. 
Many experiments~\cite{Evered_2023_gate, Bluvstein_2024_Ryd, Saffman_2025_atom} also found that the infidelity caused by crosstalk during parallel realization of quantum gates is marginal. 


\section{\label{sec:Grover}II.\quad Grover's search algorithm}
Grover's search algorithm~\cite{Grover} is an algorithm based on amplitude amplification~\cite{Nielsen} to solve NP problems. It is a black-box algorithm that requires an oracle corresponding to the target NP problem. For an NP problem with $n$ variables and function $f:\{0,1\}^{\otimes n} \mapsto  \{0,1\}$, the Grover oracle $U_w$ is required to realize a $\pi$-phase shift  for the legal (solution) states with respect to the illegal (non-solution) states: 
\begin{equation}
    U_w:|z\rangle \to (-1)^{f(z)}|z\rangle .
    \label{O}
\end{equation}
Define the superposition state of all possible solutions as $|\psi \rangle \equiv \sum_{z=0}^{2^n-1}|z\rangle$ and the Grover diffusion operator $U_s:=2|\psi \rangle \langle \psi|-I = (HX)^{\otimes n} (C^{n-1}Z) (XH)^{\otimes n}$, where $C^{n-1}Z$ represents a multi-qubit Z gate and $H,X$ are single-qubit gates. Applying the Grover operator $G=U_sU_w$ (Fig.~\ref{fig:G}) on the superposition state repeatedly can amplify the amplitude of the solutions. 
It can be proved that after $O(2^{n/2})$ iterations, the probability of obtaining a solution in the measurement is greater than 0.5~\cite{Nielsen}. 
\begin{figure}[htb]
  \includegraphics[width=0.35\textwidth]{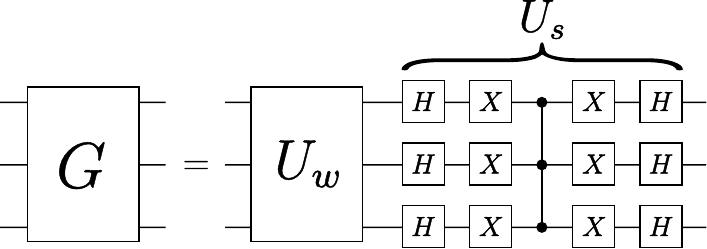}
  \caption{\label{fig:G} Grover operator G in Grover's search. It is able to amplify the amplitude of the solutions. }
\end{figure}

Since the number of rotations required is of order $O(2^{n/2})$, there is an explicit quadratic speedup compared to classical search algorithms, whose complexity for solving an NP problem with $n$ variables is $O(2^n)$ under the strong exponential-time hypothesis~\cite{2001_SETH,2001_kSAT}. 
Regarding circuit depth, the multi-controlled Z gate in $U_s$ can be realized using the QBT introduced in the main text and achieves a depth of $O(\mathrm{log}_2n)$. 
{
Therefore, if the depth of the oracle is $O(d(n))$, the overall complexity is $O\left((d(n)+\mathrm{log}_2n)\cdot 2^{n/2}\right)$. }

\section{\label{sec:checking}III.\quad  The checking circuit and its transpilation}
As shown above, the key to constructing a quantum solver for NP problems lies in the realization of the Grover oracle. Therefore, we introduce the unified framework applicable to a broad class of NP problems, encompassing Karp's 21 NP-complete problem~\cite{Karp:NPC}, to construct quantum oracles and realize them on the Rydberg atom system. 

In this section, we will discuss the specific construction of the checking circuit in the framework in detail. We rewrite the unified framework for the decision version of NP problems: 
\begin{equation}
    f(z)=\bigwedge_{\mu=1}^{p} g_\mu({\cal A}_\mu) \wedge H\left(\sum_{\nu=1}^{q} h_\nu({\cal B}_\nu)-k\right), 
\label{eq:f(x)}
\end{equation}
where ${\cal A}_\mu =\{z_i| i \in {\cal I}^A_\mu\},{\cal B}_\nu =\{z_i| i \in {\cal I}^B_\nu\}$ and $z_i$ represent the variables (see in the main text). The function $H(x)$ is the Heaviside function, whose specific value will be discussed later. 
In general, the $H$ function in the equation can appear more than once. However, since only the Knapsack problem falls into this category and it can be easily addressed by adding an extra $M^{[2]}$ in the merging circuit, we omit the possible multiplexing here for simplicity. 

Among the Karp's 21 NP-complete problems~\cite{Karp:NPC}, we select some representative problems to specify the constraints $g_\mu$ ($h_\nu$) for them: the satisfiability problem (SAT), the set cover problem (SCP), the clique cover problem (CCP), the node cover problem (NCP), the hitting set problem (HSP), the clique problem (CQP), the exact cover problem (ECP), the max cut problem (MCP), the Knapsack problem (KSP) and the undirected Hamiltonian cycle problem (HCP). The unit construction for the rest of the NP-complete problems resembles these representative problems. 
There are more NP problems studied before, among which we choose the most famous and widely discussed ones, the maximum independent set problem (MIS)~\cite{Ebadi_2022_QA, Tibaldi_2025_QAOA}, the dominant set problem (DSP)~\cite{Jiang_2023_DSP} and the number partition problem (NPP)~\cite{Anikeeva_2021_NPP}. 
The constraint functions of these NP problems are shown in Tab.~\ref{tab:NPC}. 
\begin{table}[tp]
\centering
\begin{tblr}{
  colspec = { 
  Q[c,wd=2.5cm]
  Q[c,wd=0.5cm]
  Q[c,wd=1cm]
  Q[c,wd=1.5cm]
  Q[c,wd=2.5cm]
  Q[c,wd=2cm]
  Q[c,wd=2cm]
  },
  row{1} = {rowsep=0pt, abovesep = 3pt, belowsep = 1pt}, 
  row{2,13} = {rowsep=0pt, abovesep = 3.5pt, belowsep = 0pt},
  row{3-11} = {rowsep=0pt, abovesep = 0.5pt, belowsep = 0pt}, 
  row{12,14} = {rowsep=0pt, abovesep = 1pt, belowsep = 2pt}, 
  hline{1, Z} = {1pt},
  hline{2, 13} = {0.5pt},
  stretch = 0.9,
}
 Problem& $\tilde{n}$& $N$& $t$& $g_\mu$ & $h_\nu$ & Note \\ 
 $(n, m, k)$ SAT& $n$& $m$& $k$& $x_{i_{\mu 1}}\vee... \vee \neg x_{i_{\mu k}}$& & \\ 
 $(n, v, d,k_1)$ SCP& $n$& $(v,n)$& $(d,1)$& $s_{i_{\mu 1}}\vee... \vee s_{i_{\mu d}}$& $s_\nu$& \\ 
 $(n, e, k_1)$ NCP& $n$& $(e, n)$& $(2,1)$& $v_{i_{\mu 1}}\vee v_{i_{\mu 2}}$& $v_\nu$& \\ 
 $(n, m, k, k_1)$ HSP& $n$& $(m, n)$& $(k, 1)$& $v_{i_{\mu 1}}\vee ... \vee  v_{i_{\mu k}}$ & $v_\nu$& \\ 
 $(n, \overline{e}, k_1)$ CQP& $n$& $(\overline{e},n)$& $(2,1)$& $\neg(v_{i_{\mu 1}}\wedge v_{i_{\mu 2}})$& $v_\nu$& \\ 
 $(n, v, d)$ ECP& $n$& $v$& $d$& $\delta(\sum_{j=1}^d s_{i_{\mu j}}-1)$& & \\ 
 $(n, e, k_1)$ MCP& $n$& $e$& $2$& & $v_{i_{\nu 1}}\oplus v_{i_{\nu 2}}$& \\ 
 $(n, k_1, k_2)$ KSP& $n$& $(n,n)$& $(1,1)$& $v_\nu \cdot wgt_\nu \ ^*$ & $v_\nu \cdot val_\nu$ &$^*$ Also $h_\nu$ \\ 
 $(n, e, k_1)$ MIS& $n$& $(e,n)$& $(2,1)$& $\neg(v_{i_{\mu 1}}\wedge v_{i_{\mu 2}})$ & $v_\nu$& \\ 
 $(n, d, k_1)$ DSP& $n$& $(n, n)$& $(d+1,1)$& $v_{i_{\mu 0}}\vee ... \vee  v_{i_{\mu d}}$ & $v_\nu$& \\ 
 $(n, k_1)$ NPP& $n$& $n$& $1$& & $v_\nu \cdot val_\nu$ & \\ 
 $(n, \overline{e}, k_1)$ CCP& $mn$& $\overline{e}$& $2m$& 
 $\neg \delta(o_{i_{\mu 1}}-o_{i_{\mu 2}})$
 & &$m:=\left \lceil\mathrm{log}_2k_1 \right \rceil$  \\ 
 $(n, e, d)$ HCP$ ^*$& $e$& $n$& $d$& $\delta(\sum_{j=1}^d e_{i_{\mu j}}-2)$ & & $ ^*$ Special \\ 
\end{tblr}
\caption{\label{tab:NPC} Constraints for NP problems. Different variable notations are used here to reflect their distinct semantic roles in each problem. 
Definitions of all parameters are provided in the main text. The critical size, appearing in the decision version of some NP problems to constrain the size of the solutions, are denoted with $k_1$ (and an extra $k_2$ for KSP). The first and second element of $N,t$ are for $g_\mu$ and $h_\nu$, respectively. 
SAT: $x_{i_{\mu j}}$ denotes the $j$-th variable in the $\mu$-th clause.
SCP/ESP: $s_{i_{\mu j}}$ denotes the $j$-th subset that contains the $\mu$-th element. 
NCP/CQP/MCP/MIS: $v_{i_{\mu j}}$ denotes the $j$-th vertex connected to the $\mu$-th edge. 
HSP/KSP/NPP: $v_{i_{\mu j}}$ denotes the $j$-th element in the $\mu$-th set. 
CCP: $o_{i_{\mu j}}$ denotes the subclique index of the $j$-th vertex connected to the $\mu$-th edge. $m=\left \lceil\mathrm{log}_2k_1 \right \rceil$ is a constant irrelevant to problem size $n$. 
DSP: $v_{i_{\mu j}}$ denotes the $j$-th vertex connected to the $\mu$-th vertex. $v_{i_{\mu 0}}$ denote the $\mu$-th vertex itself. 
HCP: $e_{i_{\mu j}}$ denotes the $j$-th edge connected to the $\mu$-th vertex. 
The integers $wgt_\nu, val_\nu$ in KSP/NPP are the weight, value of the $\nu$-th item. For KSP/NPP, $b$ for $h_\nu$ is determined by the largest integer in $wgt_\nu$ and $val_\nu$, while for other problems $b=1$. For SAT, the constraint shown here is a representative example. There may be an $\neg$ operation before arbitrary variables. For HCP, a special structure is needed, which will be shown later. For $h_\nu=v_\nu$ or $s_\nu$, the checking units are just identity and thus can be omitted as in the main text. $\delta(x)$ here means $\delta_{x,0}$. }
\end{table}
The definitions of the parameters in the NP problems in Tab.~\ref{tab:NPC} are summarized as follows: 
\begin{equation}
    \begin{cases}
        \text{SAT: $n$ variables, $m$ clauses, $k$ variables in each clause } \\
        \text{HSP: $n$ elements, $m$ subsets, $k$ elements in each subset } \\
        \text{DSP: $n$ vertices, $d$ vertices connected to one chosen vertex } \\
        \text{SCP\&ECP: $n$ subsets, $v$ elements, $d$ subsets that contain one chosen element } \\
        \text{CQP\&CCP\&NCP\&MCP\&MIS: $n$ vertices, $e(\overline{e})$ edges in the (complement) graph } \\
        \text{HCP: $n$ vertices, $e$ edges, $d$ edges connected to one chosen vertex } \\
        \text{KSP\&NPP: $n$ elements} .
    \end{cases}
\end{equation}
following the mostly accepted definitions~\cite{Karp:NPC, Garey_1990_NPC, Moore_2015_d_regular}. 
We use $k_1$ ($k_2$) to represent a critical size of an NP problem~\cite{Garey_1990_NPC, Goldreich_2010_NPC} that can constrain the size of solutions. The specific question of the problem determines the form of the $H$ function in Eq.~\ref{eq:f(x)}. Specifically, for questions asking whether there exists a solution with size strictly larger (smaller) than $k_1$, the function is $H(x)=\theta(x)$ ($\theta(-x)$) and satisfies $H(0)=0$. 
Here, the problem instances are assumed to be uniform, such as $k$-uniform SAT~\cite{Karp:NPC}, or regular, such as $d$-regular ECP~\cite{Moore_2015_d_regular}. We will discuss the non-uniform or non-regular problem instances later. 

This solver has three key parameters, $(\tilde{n},N,t)$ in Tab.~\ref{tab:NPC}, representing the number of (data qubits, checking units, data qubits needed in each checking unit). These parameters, taking different values for different NP problems, play a crucial role in estimating the cost of the solver. 
For problems with checking units that have more than one form, $N$ and $t$ become vectors. However, note that $h_\nu=v_\nu$ or $s_\nu$ are directly the form of data qubits and the corresponding checking units are identity. Therefore, for concise expression, they can be omitted as in Tab.~1 in the main text. 
Then, all checking units in a given NP problem are identical and $N,t$ become scalars. Compared to the definitions in Eq.~\ref{eq:f(x)}, for a problem that only has $g_\mu$ ($h_\nu$), the parameter $N$ is equal to $p$ ($q$). 
Note that although the units that are identity are omitted, the merging circuit $M^{[2]}$ for them still contributes to the depth. 
The number of data qubits $\tilde{n}$ is equal to the number of variables. For most NP problems, the number of variables is directly defined as $n$ in their standard definitions. For some rare exceptions, such as CCP and HCP, the notation $n$ has a specific meaning in the original definition, as shown in Tab.~\ref{tab:NPC}. 
For simplicity, we henceforth use $n$ to denote the number of data qubits in the following, and address any exceptions individually if necessary.

While the logic behind most constraints in Tab.~\ref{tab:NPC} is relatively straightforward, some problems—such as CCP, DSP, and HCP—require further clarification due to their more implicit structural requirements. We elaborate on these cases below. 
In CCP, the $n$ vertices are each assigned $m$ qubits to encode the index (from $1$ to $k_1$) of the subclique they belong to. For every edge in the complement graph, the two vertices connected to it should not be in the same subclique or the configuration is illegal. 
In $d$-regular DSP, for the $\mu$-th vertex $v_{i_{\mu 0}}$, only when itself and its nearest neighbors $v_{i_{\mu j}}, j=1,...,d$ are not chosen (at $|0\rangle$) will a corresponding configuration be illegal. 

In HCP, $g_\mu$s in the table make sure that for any vertex, there are exactly two edges connecting to it. 
However, these constraints alone are insufficient to guarantee that the selected edges form a single Hamiltonian cycle; the candidate configuration may consist of multiple disjoint cycles. 
To implement this connectivity check, we assign an activation qubit (initialized to $|0\rangle$) to each vertex. The process begins by setting one of these activation qubits to $|1\rangle$ (representing the start vertex) and proceeds through $\lfloor n/2 \rfloor$ rounds of activation propagation. In each round, all $e$ edges are traversed: if a vertex is activated and the edge it connects to is included in the candidate solution, the activation is propagated to the adjacent vertex in the next level. 
The detailed process is shown in Fig.~\ref{fig:Connected}. 
\begin{figure}[tp]
    \begin{center}
  \includegraphics[width=0.48\linewidth]{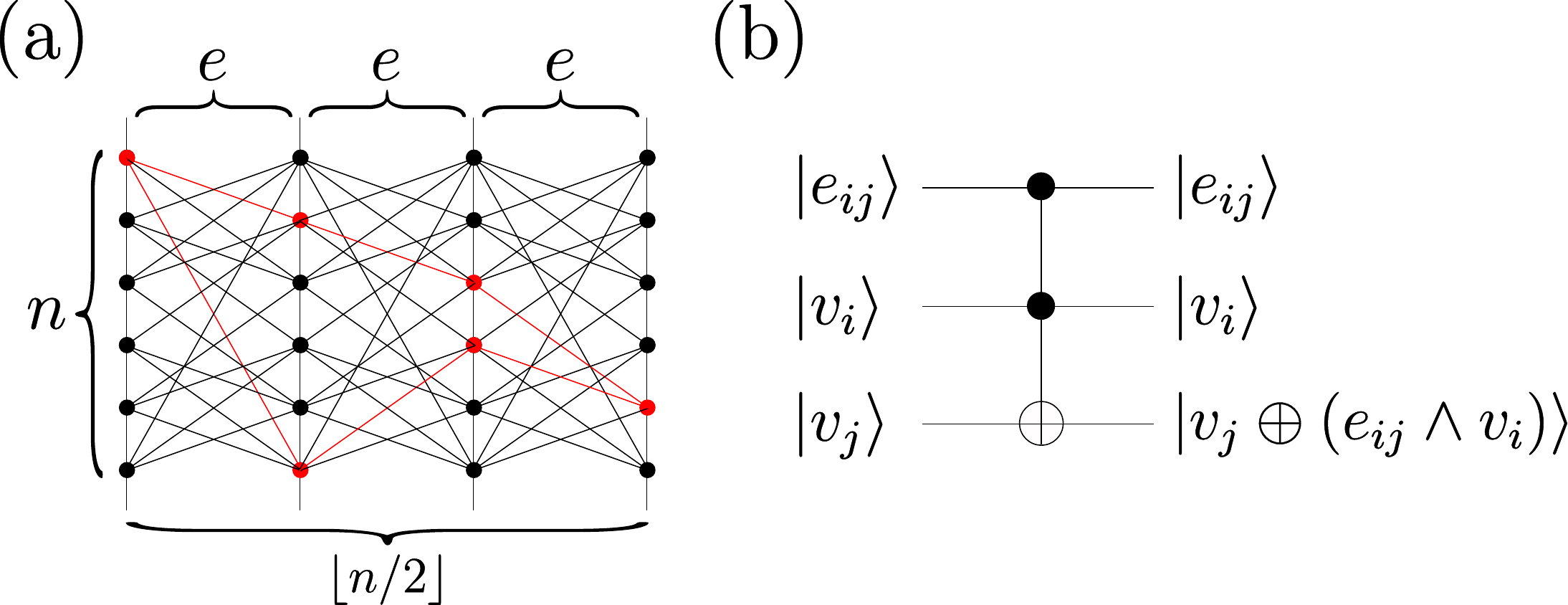}
    \end{center}
  \caption{\label{fig:Connected} The special structure for HCP to check for the connectivity of any possible solution. Note that the original variables in HCP represent the edges in Tab~\ref{tab:NPC}, which already exist. 
  (a) The checking process. The $n$ points in each level represent the activation qubits of $n$ vertices in a graph. Starting from one chosen vertex (set to $|1\rangle$) and, in each level of activation, all $e$ edges are traversed to try to activate the vertices in the next level. 
  To activate all $n$ vertices in a possible Hamiltonian cycle, $\lfloor n/2 \rfloor$ levels are needed. A valid Hamiltonian cycle is shown in red edges and vertices for example. 
  (b) The checking unit for activation corresponding to an edge $e_{ij}$. The vertices connected to it are $v_i,v_j$. Here $|v_i\rangle$ ($|v_j\rangle$) is the activation qubit for vertex $v_i$ ($v_j$). The vertex $v_j$ is in the next level of vertex $v_i$. }
\end{figure}
At the end of the propagation process, a valid Hamiltonian cycle will result in all activation qubits being set to $|1\rangle$ if $n$ is odd, or all but one if $n$ is even. For disconnected cycles, at least three activation qubits remain in the $|0\rangle$ state.
These qubits are independent of those used in Tab.~\ref{tab:NPC} and should be merged by $M^{[2]}$ to check whether the number of activated qubits (at $|1\rangle$) is $n$ ($n-1$) for odd (even) $n$. 
Note that each level in Fig.~\ref{fig:Connected} (a) is the same group of checking units with $N=e$ and $t=2$ (Fig.~\ref{fig:Connected} (b)), which can be performed using the algorithm introduced in the main text. The process requires $\lfloor n/2 \rfloor$ repetitions of this group. 
Hence, taking the problem size as the number of edges $e$, the depth of this process is of order $O(e/n \cdot \lfloor n/2 \rfloor)=O(e)$ and the required qubit number is $O(e+n)$. 
At the HCP hardness threshold $e=O(n\mathrm{log_2}n)$~\cite{Komlos_1983_HCP_threshold}, the qubit number cost and the circuit depth are both $O(e)$. 
The depth of HCP is different from other problems due to this expensive special structure. Problems that require a check for graph connectivity need this structure. 

For simplicity, SAT, HSP are assumed $k$-uniform and SCP, ECP, DSP are assumed $d$-regular in previous discussions. For non-uniform or non-regular problem instances, they can be easily transformed to uniform or regular instances by adding a constant number of auxiliary subjects. We take a non-regular set cover problem with $n$ subsets and $v$ elements as an example to illustrate it. For its constraint functions, assuming that $|\mathcal{A}_\mu|=d_\mu, \mu = 1,...,v$, define 
\begin{equation}
 d_{\mathrm{min}}=\mathrm{min}_\mu\{d_\mu\}, d_{\mathrm{max}}=\mathrm{max}_\mu\{d_\mu\}. 
\end{equation}
Then add $d_{\text{aux}} := d_{\mathrm{max}}-d_{\mathrm{min}}$ auxiliary subsets $n^{\text{aux}}_i, i=1,...,d_{\text{aux}}$ to the original $n$ subsets. For each index $\mu$, setting 
\begin{equation}
    s_\mu \in n^{\text{aux}}_i, i=1,...,d_{\mathrm{max}}-d_\mu, 
\end{equation}
where $s_\mu$ represents the element corresponding to the variable set $\mathcal{A}_\mu$. Then, the instance becomes a $d_{\mathrm{max}}$-regular instance with $n+d_{\text{aux}}$ subsets. Note that the auxiliary subsets should not be in the possible solution. 
This means that the auxiliary qubits that represent them should remain in $|0\rangle$ throughout the process, which can be achieved by removing the $H$ gates during initialization. Since $d_{\text{aux}}$ is not related to the problem size, the cost remains unchanged. 

The specific circuits of the checking units can be constructed directly from the functions in Tab.~\ref{tab:NPC}, because they only consist of basic operations. Some representative checking units are constructed in Fig.~\ref{fig:Checking}. 
\begin{figure}[tp]
    \begin{center}
  \includegraphics[width=0.4\linewidth]{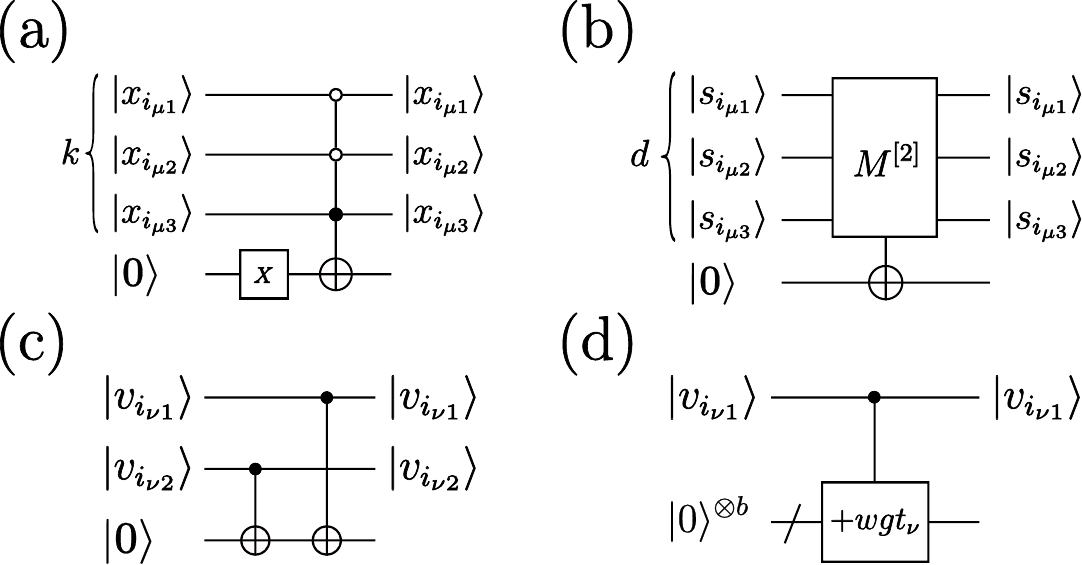}
    \end{center}
  \caption{\label{fig:Checking} Some representative checking units $C_\mu$, $D_\nu$ that realize $C_\mu|{\cal A}_\mu\rangle|0\rangle \mapsto |{\cal A}_\mu\rangle|g_\mu({\cal A}_\mu)\rangle$, $D_\nu|{\cal B}_\nu\rangle|0\rangle \mapsto |{\cal B}_\nu\rangle|h_\nu({\cal B}_\nu)\rangle$. 
  (a) SAT. This is the unit for a typical clause $(x_{i_{\mu1}}\vee x_{i_{\mu1}}\vee \neg x_{i_{\mu1}})$. In the most general form, the control qubits may be controlled on 0 or 1 depending on the instance. 
  We use the method in Ref~\cite{Nielsen} to decompose the multi-controlled X gate to basic Rydberg gates, resulting in $O(k)$ depth. $k$ is an integer irrelevant to $n$. 
  (b) ECP. An $M^{[2]}$ with $b=1$ can realize the $g_\mu$ in Tab~\ref{tab:NPC}. Note that $H(x)$ should be taken as $\delta_{x,0}$. 
  Here, we decompose it by sequentially adding $s_{i_{\mu j}}$ together using a quantum adder~\cite{Takahashi_2009_RCadder}, which results in $O(d)$ depth in terms of basic Rydberg gates. $d$ is an integer irrelevant to $n$. 
  (c) MCP. (d) KSP. A controlled adder~\cite{Takahashi_2009_RCadder} is needed. $b$ is an integer related to the problem instance. The circuit for $val_\nu$ is the same. }
\end{figure} 
Since $k,d$ are both constants irrelevant to the problem size, all the units, except that of KSP/NPP, act on a constant number of data qubits and their decomposition to basic Rydberg gates results in constant depth (see Fig.~\ref{fig:Checking}). 
For KSP, the situation is a bit different. Conventionally, in complexity theory~\cite{Garey_1990_KSP, Kellerer_2004_KSP}, the number of qubits to encode them $b=\lceil \mathrm{log}_2 (\mathrm{max}\left\{ \mathrm{max}_\nu \{wgt_\nu\},\mathrm{max}_\nu \{val_\nu\}\right\}  ) \rceil$ is considered as an independent characteristic problem size. Therefore, the depth of its checking unit, using the controlled version of the ripple-carry adder, is $O(b)$ and the final circuit depth is $O(b(\mathrm{log}_2n)^2)$. 
This extra complexity is inherent in KSP, which can be improved by using a better adder~\cite{Draper_2004_QCLA}. For NPP, the discussion is the same. 
They are different from other problems as they have a unique encoding process for classical information (weights and values). 
In KSP, there are two critical sizes, the weight limit and the target value, whose corresponding constraints can be expressed as
\begin{equation}
    f(s)=\theta\left(-\sum_{\nu} h^w_\nu(s_\nu)+k_1\right)\wedge \theta\left(\sum_{\nu} h^v_\nu(s_\nu)-k_2\right) , 
\end{equation}
where $s=(s_1,...,s_n)$ are the variables and $h^w_\nu(s_\nu)=s_\nu wgt_\nu,h^v_\nu=s_\nu val_\nu$. This requires two $M^{[2]}$s to merge the information separately.  

For the transpilation of the checking circuit, we use polynomial-time classical algorithms to find sets of checking units that have no overlapping qubits. 
Since the classical algorithms only need to be run once during the whole process instead of once per Grover iteration, their complexity can be omitted. 
Under proper mapping from the qubits to the atoms, checking units in such a set can be implemented in parallel. Therefore, by rearranging atoms between two implementations of the checking unit set, the checking circuit can be performed with maximal parallelism.  
For such a checking unit set $(C_{\mu_1}, C_{\mu_2}, \ldots, C_{\mu_O})$, with their index set denoted as $\mathcal{O}$ and the qubits involved in $C_\mu$ denoted as $q_{\mu\tau}, \tau=1,\ldots, t+1$, the proper mapping from the data qubits to the atoms for the set is straightforward by neatly arranging them in a rectangle: 
\begin{equation}
\label{eq:Mapping}
P(q_{\mu \tau}) = \left[ (t+1)\left \lfloor (\mu-1)/ \lceil \sqrt{n} \rceil \right \rfloor+\tau-1, (\mu-1)\ \mathrm{mod}\ \lceil \sqrt{n}\rceil\right] , 
\end{equation} 
This strategy returns a full rectangle with $\lceil \sqrt{n} \rceil$ rows when $O\ \mathrm{mod} \ \lceil \sqrt{n}\rceil=0$. When $O\ \mathrm{mod} \ \lceil \sqrt{n}\rceil\neq 0$, we can add some completing atoms (not involved in the computation) to fill the last column during the implementation of the units, which preserves the parallelism. 

For the $k$-SAT problem, the mapping needs extra arrangement, since the $\neg$ operation that requires extra X gates may appear before arbitrary variables, making the checking units slightly different from each other. To parallelize the possible X gates, before the $i$-th rearrangement of atoms, the checking units in the $i$-th checking unit set $\mathcal{O}_i$ are categorized into at most $k$ groups by the number of $\neg$ operations. 
For each checking unit, the corresponding literals are reordered so that the negated variables appear first. Then, the constraint functions for a checking unit with $a$ $\neg$ operations can be expressed as
\begin{equation}
    g_{\mu}= \left(\bigvee_{j=1}^a \neg x_{i_{\mu j}}\right) \vee \left(\bigvee_{j=a+1}^k x_{i_{\mu j}}\right). 
\end{equation}
Then, applying the mapping separately to each group and assigning them to different columns enables parallel application of the required $X$ gates.

\begin{figure}[tp]
    \includegraphics[width=0.6\textwidth]{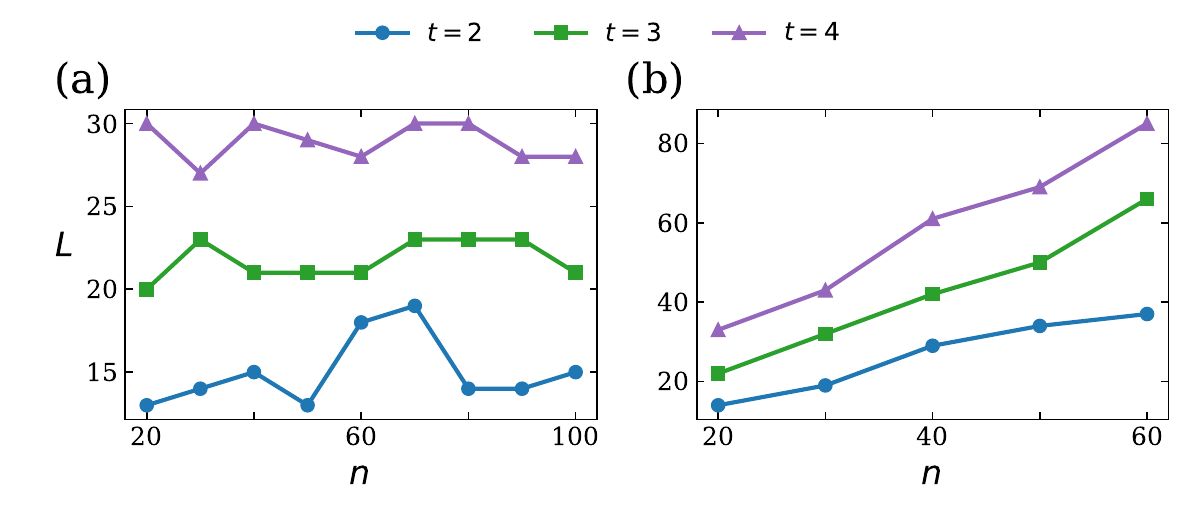}
  \caption{\label{fig:L} The number of checking unit operations $L$ with respect to $n$ for $t=2,3,4$. The hypergraphs are randomly generated with (a) $N=4n$ hyperedges, (b) $N=n^2/4$ hyperedges. The results sugguest that $L=O(tN/n)$. } 
\end{figure}

Assuming that the number of checking unit sets the algorithm find is $L$, the cost of the checking circuit is then $L$ checking unit operations since we do not include the atomic transports in the circuit depth. 
Numerical simulation in Fig.~\ref{fig:L} shows that $L$ is of order $O(tN/n)$ for a random problem instance. Since $t$ is a constant irrelevant to $n$, the relation is $L=O(N/n)$, which reduces to $O(1)$ at $N=O(n)$. 
For the high-fidelity atomic transports, they are needed between every two checking unit operations. 
Rearranging the 2D atom array to satisfy the proper map in the {\it main text} is in fact a 2D grid routing problem, which is proved to require at most $2\sqrt{n}\log n$ atomic transport~\cite{Constantinides_2024_2Droute}. 
Therefore, the total number of atomic transports is $O(\sqrt{n}\log n)$ here. 
If a stronger parallelism than the tensor-grid parallelism, selective transfer~\cite{Constantinides_2024_2Droute}, is available, this cost can be reduced to $O(\log n)$. 
Here, although the time cost scales as $O(n^{3/4}\log n)$, atomic transport is much more robust than quantum gate operations and can be redeemed as classical operations. 
In fact, atomic transports have no fundamental limit in speed and fidelity and are expected to be perfected in the near future with techniques such as magic wavelength~\cite{Jenkins_2022_magic_wavelength}.

If we consider the situation $N=O(n^2)$, the framework can be adjusted to avoid quadratic cost in space. 
A simple variant scheme can save the space cost by performing the $N=O(n^2)$ checking units separately in $\lceil N/n \rceil$ packs $\{\mathcal{O}_i\}_{i=1,2,\ldots, \lceil N/n \rceil}$ (each containing $n$ units) and restoring the ancillae between them. For simplicity, we directly use the index set $\mathcal{O}$ to represent the corresponding set of checking units. 
The scheme is shown in Fig.~\ref{fig:Variant}. 
\begin{figure}[htb]
  \includegraphics[width=0.6\linewidth]{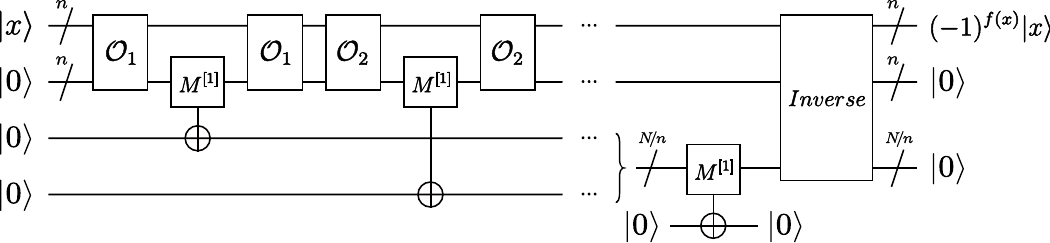}
  \caption{\label{fig:Variant} A variant scheme of the framework for situation $N=O(n^2)$. Here, the merging circuit uses $M^{[1]}$ as an example. The index sets $\mathcal{O}_i, i=1,2,\ldots, \lceil N/n\rceil$ represent the implementation of the corresponding set of checking units. Note that each checking unit is the inverse of itself. The last inverse step is used to restore the $N/n$ ancillary qubits.  } 
\end{figure}
The packs $\{\mathcal{O}_i\}_{i=1,2,\ldots, \lceil N/n \rceil}$ can still be obtained by the algorithm discussed above, except that in this situation, the size of each set is mannually upper bounded by $n$. 
The transpilation of this variant is straightforward based on that of the original framework. 
Using this variant, the qubit number cost becomes $O(2n+N/n)=O(n)$ and the depth scales as $O(n\mathrm{polylog}(n)2^{n/2})$. 

\section{\label{sec:merging}IV.\quad  The merging circuit and its transpilation}

\begin{figure}[tp]
    \begin{center}
  \includegraphics[width=0.6\linewidth]{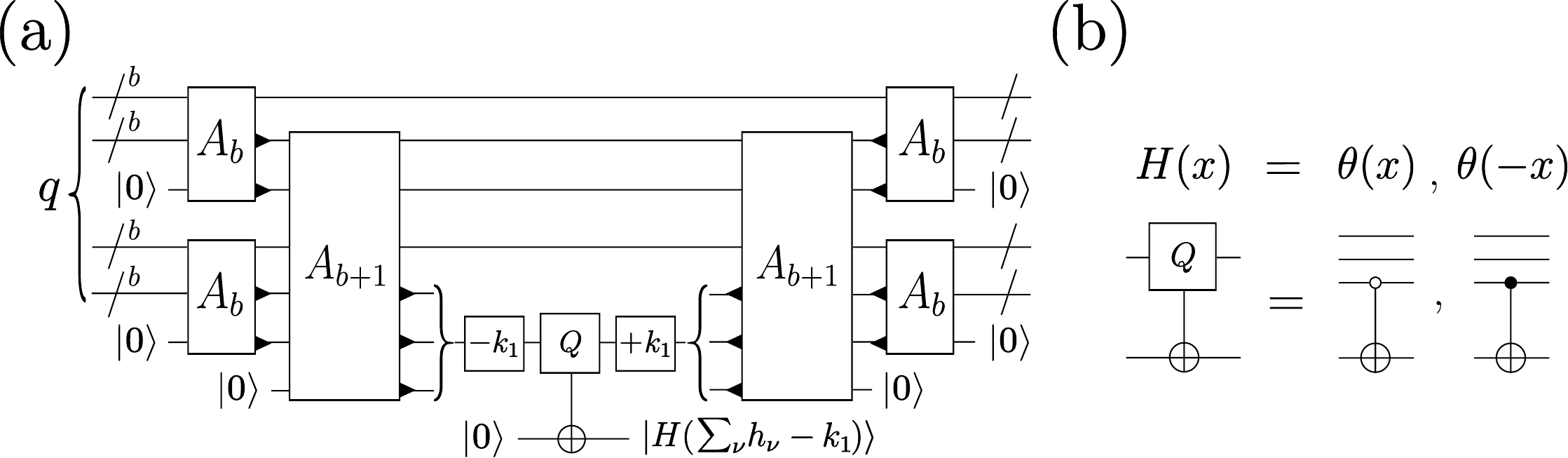}
    \end{center}
  \caption{\label{fig:QRA_cir_sup} The circuit of quantum recursive adder for $M^{[2]}$ with different $H(x)$. (a) A two-level illustration of QRA. The structure can be extended recursively. $A_u$ represents a $u$-qubit in-place adder. The ancillary qubits introduced during the circuit serve as the carry bit in these additions. (b) The Q operation in QRA. It depends on the form of $H(x)$. The lower qubit is the most significant bit. }
\end{figure}

For the merging circuit, the QBT for realizing $M^{[1]}$ is already clear in the main text. For the quantum recursive adder (QRA) for realizing $M^{[2]}$, the circuit is introduced in Fig.~\ref{fig:QRA_cir_sup}. In this work, we use the ripple-carry adder~\cite{Takahashi_2009_RCadder} as the in-place adder in QRA, which has a linear depth and does not require ancilla. 
QBT and QRA are both specifically targeted at Rydberg atom systems to achieve low cost. 
In particular, the tensor-grid qubit connectivity of the system is highly compatible with these circuits using a recursive structure, resulting in polylogarithmic depth. 

For the transpilation of them, QBT is already clear as shown in the main text. 
The transpilation of QRA is highly similar to that of QBT. In fact, we only need to replace the simple Toffoli gate in QBT with an in-place ripple-carry addition (Fig.~\ref{fig:Adder}): 
\begin{equation}
    A_u |s_1\rangle_u |s_2\rangle_u |0\rangle = |s_1\rangle_u |s_1 + s_2\rangle_{u+1},
\end{equation}
where $|s_i\rangle_u \in \mathbb{C}^{2^u}$ is a $u$-bit operand. 
\begin{figure}[bp]
  \includegraphics[width=0.5\textwidth]{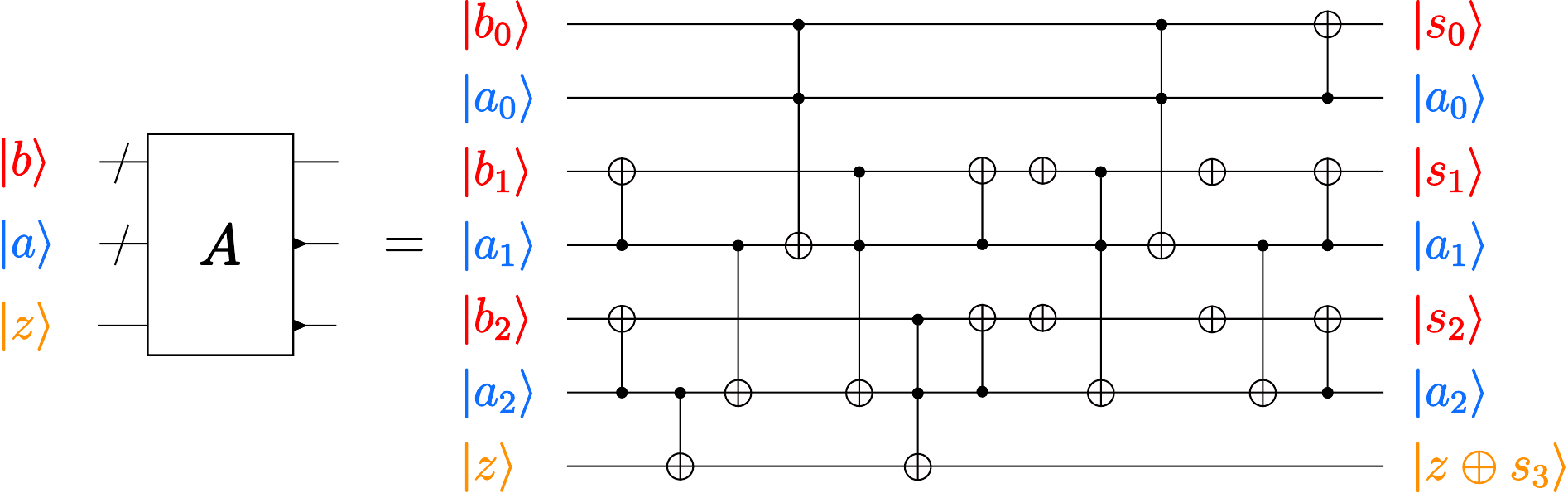}
  \caption{\label{fig:Adder} A 3-qubit ripple-carry quantum adder~\cite{Takahashi_2009_RCadder}. Red, blue, orange qubits represent the target register, the preserved operand, the carrying bit, respectively. $|s_0s_1s_2(z\oplus s_3)$ is the result. This adder is an in-place adder whose depth scales linearly with the qubit number of the operand and does not require ancilla. }
\end{figure}
Following the checking circuit, the results of $h_\nu$ are assumed to be stored in ancilla qubits arranged in a rectangular grid with height $(2b*h)$ and width $w$. 
They are separated into $h$ submodules to perform recursive additions simultaneously, as shown in Fig.~\ref{fig:QRA_sup} (a). All additions are performed along the vertical axis and the carrying bit is the lower qubit, which matches with the ripple-carry adder (Fig.~\ref{fig:Adder}). 
Specially for QRA, between each level of additions, the atoms in the target register and carrying bit of one adder in the former level need to be swapped to the preserved operand of one adder in the next level, as shown in Fig.~\ref{fig:QRA_sup} (b), in order to perform the addition in the next level. It is straightforward that all the adders in the same level can be performed in parallel due to the neat arrangement of these operands, similar to QBT. 
\begin{figure}[tp]
 \includegraphics[width=0.7\linewidth]{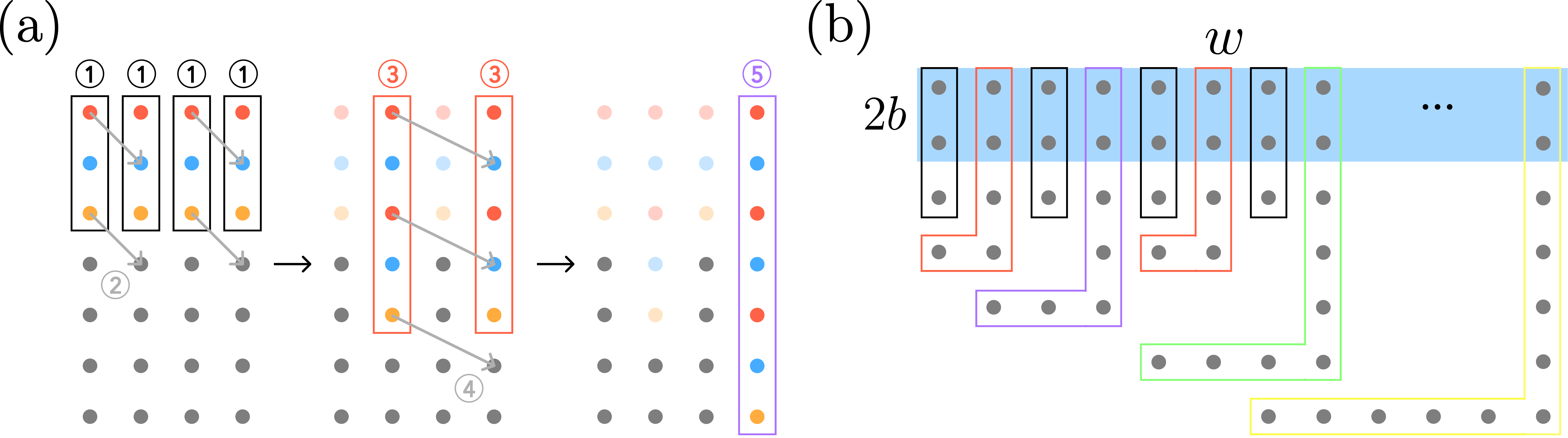}
  \caption{\label{fig:QRA_sup} Transpilation of QRA to the Rydberg atom system. The general picture is the same as QBT, while an extra step to transit between levels is needed. 
  (a) Transition between different levels of additions in QRA. Boxes with different colors contain additions in different levels. Red, blue and orange atoms here match the positions in the adder (Fig.~\ref{fig:Adder}). Between levels, parallel swaps are needed to move the output qubits of the former level to the input position of the latter level. 
  (b) General atom arrangement of QRA. 
  The blue region with shape $2b\times w$ contains part of the output of the checking circuit. 
  Boxes with different colors still contain additions in different levels. The long operands are folded to save space. 
  A submodule like this can obtain the result of the operands in the $2b\times w$ region. 
  We can arrange $h$ such submodules along the vertical axis to obtain the result of all the output of the checking circuit. This is to make the total region shape like a square instead of a long stick. Note that additions in different submodules are still in parallel. }
\end{figure}

Assuming that the checking circuit yields $N$ outputs ($b=1$), the costs of these structures in the Rydberg atom system are
\begin{equation}
    \text{QBT}
    \begin{cases}
        \text{qubits}:2N \\
        \text{depth}:4\operatorname{log}_2N
    \end{cases}, \ \ \ 
    \text{QRA}
    \begin{cases}
        \text{qubits}:2N \\   
        \text{depth}:8(\mathrm{log}_2N)^2
    \end{cases}, 
\end{equation}
where only the most significant term is reserved. 
Note that the depth here is estimated using basic Rydberg gates. 
For $N=O(n)$, these structures can be applied directly to the full output of the checking circuit.  
For $N=O(n^2)$, the structures are implemented repetitively on $n$ outputs in the variant scheme as mentioned above. 

In the former discussions, QBT starts with a $h_1\times w$ rectangle (main text) and QRA starts with a $(2b*h)\times w$ rectangle. In fact, the outputs of the checking circuit may not be a perfect rectangle. 
Here, we show that the outputs can always be transformed to what the merging circuit requires with negligible overhead. 
In general, using the mapping in Eq.~\ref{eq:Mapping}, the outputs of the checking circuit is $O(N/\lceil \sqrt{n} \rceil)$ columns, whose height is equal to or smaller than $\lceil \sqrt{n} \rceil$. 
Then, by stacking the columns whose height is smaller than $\lceil \sqrt{n} \rceil$ to form complete columns with height $\lceil \sqrt{n} \rceil$, the full output of the checking circuit can be arranged into a complete rectangle with $\lceil \sqrt{n} \rceil$ rows. 
Since one column can be moved at once, the stacking process at most requires $O(N/\lceil \sqrt{n} \rceil)$ steps of atomic transport. 
It is reasonable to take $h$ ($h_1$) as $z*\lceil \sqrt{n} \rceil, z\in \mathbb{N}^+$ while merging. Under this choice, approximately $z$ more transports are needed to rearrange the full output of the checking circuit to the expected shape. 
Note that the number of atomic transports in the checking circuit is of order $O(N)$. Therefore, the total number of extra transports here is always negligible. 

For the Grover diffusion operator $U_s$ (Fig.~\ref{fig:G}), it is implemented directly on the data qubits. The single-qubit gates in it are obviously parallelizable. The multi-controlled Z gate in it can be realized using QBT. 
{
Finally, at $N=O(n)$, for an NP problem that does not have $h_\nu$, the circuit depth of the Rydberg atomic solver is of order $O(\mathrm{log}_2N \cdot 2^{n/2})$. For an NP problem having $h_\nu$, the circuit depth is of order $O((\mathrm{log}_2N)^2 \cdot 2^{n/2})$. The required qubit number is of order $O(n)$. }

\section{V.\quad Compatibility with QEC codes}
Our protocol includes basic Rydberg gate operations and atomic transport in the Rydberg atom system. Although the state-of-the-art performances of their fidelity~\cite{Evered_2023_gate, Bluvstein_2024_Ryd} are extraordinary, for fault-tolerant quantum computing with large system size, quantum error correction (QEC) is necessary. 
In this section, we briefly discuss the compatibility of our protocol with some QEC codes. 

For some CSS codes, such as Shor code $[[9,1,3]]$~\cite{Shor_1996_QEC} or Steane code $[[7,1,3]]$~\cite{Steane_1997_QEC}, they use certain physical qubits to encode one logical qubit and can be scaled up to more logical qubits by generating copies. 
In our protocol, applying these codes is to replace each atom with a small group of atoms, which can be neatly arranged in a line or a rectangle. 
In current QEC community, an important framework is to use transversal gates combined with high-fidelity magic states~\cite{Rodriguez_2024_Magic_state} for universal quantum computing to prevent error proliferation. 
A transversal gate is a logical operation implemented by applying physical gates independently across code blocks, such that each physical gate acts only on corresponding qubits. 
Note that the property---physical gates acting on corresponding qubits in each block---is the same as the situation in the checking circuit (see Fig.~2 in the main text), which is compatible with the tensor-grid pattern parallelism. Thus, the physical gates corresponding to some originally parallel logical gates are still parallelizable in Rydberg atom systems. Likewise, parallel atomic transport is also preserved. 
Therefore, no parallelism in our protocol is lost when applying these CSS codes and the cost estimate at the complexity level holds. Note that this statement still holds when concatenation is used for larger code distances. 

For topological codes~\cite{Nye_2024_topological_QEC} or LDPC code~\cite{Tillich_2014_LDPC, panteleev_2022_LDPC, Pecorari_2025_LDPC}, if we still use one code block for each logical qubit and scale up by generating copies and concatenation, the statement above still holds. 
If we directly compute in the large and complex code space of a topological code, some gate operations may be very complicated. The compatibility with such codes is beyond the scope of this work. 

In our protocol, the atomic transport takes a long time. With a reasonable moving speed, this process comes essentially with no fidelity cost~\cite{Bluvstein_2024_Ryd}. 
If we increase the moving speed to reduce time, the atoms may heat up a bit which eventually causes atom loss error. Fortunately, the atom loss error in Rydberg atom systems can also be corrected by QEC as proposed by recent works~\cite{Baranes_2025_atom_loss, Bluvstein_2025_atom_loss}. 

In summary, our protocol is compatible with many QEC codes and can be useful in future fault-tolerant quantum computing. 

\section{VI.\quad Possible demo for one Grover iteration}
One Grover iteration of the atomic solver can be demoed on a state-of-the-art Rydberg atom platform for small problem sizes. 

We take the 3-SAT problem with $m=n$ as an example and carefully count the number of quantum gates and atomic transports in this circuit. The result is shown in Tab.~\ref{tab:demo}. 
The checking unit of this problem is a CCCX gate (where certain control qubits may be controlled on 0). For its decomposition, we use the simple method in Ref~\cite{Nielsen}. 
For $n=8$, this protocol uses 24 qubits, 74 CCZ gates (46 in depth) and 212 H/X gates (46 in depth) along with at most 57 atomic transports. Suppose that the moving speed is $0.5\mu m/\mu s$, which is reported to have no fidelity loss~\cite{Bluvstein_2024_Ryd}, the time required for all transports is about hundreds of microseconds and the time for gate operations is about one hundred microseconds. 
The total time does not exceed the lifetime of the cold atoms in a well-built system~\cite{Evered_2023_gate}. 

\begin{table}[tp]
\centering
\begin{tblr}{
  colspec = { *{8}{Q[c,wd=0.72cm]} },
  row{1} = {rowsep=0pt, abovesep = 3pt, belowsep = 1pt}, 
  row{2} = {rowsep=0pt, abovesep = 3pt, belowsep = 1pt},
  row{3-5} = {rowsep=0pt, abovesep = 3pt, belowsep = 0pt}, 
  hline{1, Z} = {1pt},
  hline{2, 3} = {0.5pt},
  stretch = 0.9,
}
\SetCell[r=2]{c} $n$
  & \SetCell[r=2]{c} $m$
  & \SetCell[r=2]{c} Qubit
  & \SetCell[c=2]{c} Gate &  
  & \SetCell[c=2]{c} Depth &   
  & \SetCell[r=2]{c} TP 
  \\
& & & CCZ & SQ & CCZ & SQ & 
\\
8  & 8  & 24  & 74  & 212 & 46 & 46 & 57 \\
16 & 16 & 48  & 154 & 436 & 50 & 50 & 112 \\
64 & 64 & 192 & 634 & 1780 & 58 & 58 & 352 \\
\end{tblr}
\caption{\label{tab:demo} 
The required resources for one Grover iteration of the atomic solver for the 3-SAT problem at $n=8,16,64$. 
For simplicity, we take $m=n$ here. SQ means the number of single-qubit gates and TP means the number of atomic transports. 
Here, the single-qubit gates contain H and X gates only. H gate is required in the decomposition from CCX to CCZ and X gate is required as some CCX may be controlled on 0. They are also needed in the diffusion operator.}
\end{table}

In 2024, an important work~\cite{Bluvstein_2024_Ryd} in atomic quantum computing has realized 228 logical two-qubit gates and 48 logical CCZ gates in a system with 48 logical qubits. 
Therefore, it is entirely feasible for a small demo of our protocol for $n=8$ to be realized on current Rydberg atom platforms. 
In the NISQ era, the resources for $n=16, 64$ are also not too formidable.

\section{\label{sec:sc}VII.\quad  Comparison to the superconducting platform}
In this section, we briefly discuss the transpilation overhead of this framework to a standard superconducting chip with fixed local qubit connectivity~\cite{Arute_2019_Google, AbuGhanem_2025_IBM} and compare it to the Rydberg atom platform. 

Similarly to the Rydberg atom system, the $n$ data qubits in an NP problem are mapped to a 2D superconducting qubit array with side lengths of order $\sqrt{n}$. 
On a standard superconducting chip with fixed local qubit connectivity, the entangling gates between the nearest-neighboring atoms can be directly implemented, while the remote gates rely on swap gates to bring the involving qubits into proximity. 
For a random NP problem instance, the checking units are in general non-local, which means that the entangling gates in them may involve qubits that are far away from each other. 
The number of non-local gates is of order $O(N)$ and the distance between two qubits in a non-local gate is on average $O(n^{1/2})$. 
Therefore, even with maximal parallelism (which may not be possible for many technical reasons), the total depth of swaps in the checking circuit is $O(N/n\cdot n^{1/2})$. 
For comparison, the circuit depth on the Rydberg atom platforms is $O(N/n)$, offering a significant speedup. 

For the merging circuit, the fixed local qubit connectivity of the superconducting chip does not support the parallelism of the binary tree structure. 
In fact, assuming that the number of control qubits is $N$, the result of the previous work~\cite{Chu_2023_sc_gate} shows that the depth of this multi-controlled X gate ($M^{[1]}$) on a 2D superconducting qubit array is of order $O(\sqrt{N})$. 
For comparison, the depth of $M^{[1]}$ on the Rydberg atom platforms is of order $O(\mathrm{log}_2N)$, offering an exponential speedup. The discussion for $M^{[2]}$ is similar. 

Generally, for a Grover oracle, assuming $N=O(n)$, the circuit depth in a Rydberg atom system is of order $O(\mathrm{polylog(n)})$ while the circuit depth on a standard superconducting chip with fixed local qubit connectivity is of order $O(n^{1/2})$. To be more specific, we can estimate the resources for small problem sizes using the problem setting in the former section. For $n=16$, the required entangling gates (depth) are 125 CCZs (43) in Rydberg atom systems and at least 896 CCZ/CZs (188) in superconducting platforms. For $n=128$, the required entangling gates (depth) are 1021 CCZs (49) in Rydberg atom systems and at least 19456 CCZ/CZs (492) in superconducting platforms. Note that this estimation have omitted the overhead for decomposing CCZ gates in superconducting platforms and the classical atomic transports (perfect fidelity) needed in Rydberg atom systems. 



\end{widetext}

\end{document}